\documentclass[12pt]{article}
\usepackage{amsmath}

\numberwithin{equation}{section}
\numberwithin{figure}{section}
\numberwithin{table}{section}

\usepackage{color}
\usepackage{caption}
\usepackage{subcaption}
\usepackage{graphicx}
\usepackage{enumerate}
\usepackage{natbib}
\usepackage{url}
\usepackage[hidelinks]{hyperref} 
\usepackage{booktabs}
\usepackage{stackengine}

\usepackage{xcolor}
\usepackage{subcaption}
\DeclareCaptionFormat{custom}
{
    \textbf{#1#2}\textit{\small #3}
}
\newcommand{\blind}{1}

\usepackage[normalem]{ulem}
\usepackage{soul}

\usepackage{xcolor}
\definecolor{dpink}{rgb}{1,0,1}

\definecolor{hypcol}{rgb}{0.5,0,1}
\hypersetup{colorlinks=true,linkcolor=hypcol,anchorcolor=hypcol,citecolor=hypcol,filecolor=hypcol,urlcolor=hypcol,bookmarksnumbered=true}

\newcommand{\MSE}{\text{MSE}}
\usepackage{float}
\newcommand{\MSD}{\text{MSD}}

\begin{document}

\def\spacingset#1{\renewcommand{\baselinestretch}%
{#1}\small\normalsize} \spacingset{1}

%%%%%%%%%%%%%%%%%%%%%%%%%%%%%%%%%%%%%%%%%%%%%%%%%%%%%%%%%%%%%%%%%%%%%%%%%%%%%%

\if1\blind
{
  \title{\bf Improving measurement error and representativeness in nonprobability surveys}
  \author{Aditi Sen and Partha Lahiri \thanks{
Aditi Sen is PhD Candidate, Applied Mathematics \& Statistics, and Scientific Computation (e-mail: asen123@umd.edu) and Partha Lahiri is Professor, Department of Mathematics and The Joint Program in Survey Methodology, University of Maryland, College Park, MD  20742, USA (e-mail: plahiri@umd.edu).}\hspace{.2cm}}
\date{}
  \maketitle
} \fi 

\if0\blind
{
  \bigskip
  \bigskip
  \bigskip
  \begin{center}
    {\LARGE\bf Improving measurement error and representativeness in nonprobability surveys}
\end{center}
  \medskip
} \fi

\bigskip

\begin{abstract}

In the age of big data, nonprobability surveys are becoming increasingly abundant. Data integration techniques involving both probability and nonprobability surveys are being extensively used for providing improved estimates for finite population estimation. While much of the existing research has focused on mitigating selection bias in nonprobability surveys, the issue of measurement error within these surveys remains relatively unexplored. Statistical methods devised with the purpose of reducing selection bias are appropriate for reliable estimation, only under the assumption of accuracy of survey responses. Motivated by a recent case study of \cite*{KML}, our research addresses bias from both measurement and sampling errors in nonprobability surveys. In this article, we propose a new data integration method that uses multiple probability and nonprobability surveys and leverages machine learning models to construct a composite estimator. The proposed composite estimator integrates probability and nonprobability surveys, when both contain response variables of interest. We analyze the performance of this estimator in comparison to an existing composite estimator in literature, analytically as well as empirically, using multiple survey data from \cite{KML}. Finally, we identify conditions under which the proposed estimator outperforms estimators based solely on probability surveys.

\end{abstract}

\noindent%

\spacingset{1.45} % DON'T change the spacing!

\newpage

%%%%%%%%%%%%%%%%%%%%%%%%%%%%%%%%%%%%%%%%%%%%%%%%%%%%%%%%%%%%%%%

\section{Introduction}
\label{sec_intro}

The use of design-based inference from probability samples has been pivotal for producing reliable estimates for finite population characteristics (such as means, totals, etc.) ever since the groundbreaking research of \cite{Neyman1934}. It is attractive in the sense that the same estimation methodology can be applied to any response variable of interest without the need of explicit modeling when the following are available: a sampling frame (list of all units in the targeted finite population), a sampling design with known and nonzero selection probabilities and the survey responses {from the selected sample.} 
Despite the presence of unequal selection probabilities, design-based inference remains valid in the case of probability samples; for example, the Horvitz–Thompson estimator (\cite{horvitzthompson}) yields unbiased estimates by incorporating the known inclusion probabilities. In contrast, nonprobability samples present a fundamentally different challenge: the selection mechanism is unknown, and some population units may have zero probability of selection. This absence of design information renders traditional design-based methods inapplicable and motivates the development of alternative inferential strategies.

Even though it is desirable to produce estimates based on probability samples, it may not always be practically possible due to budgetary and other constraints. It can be very expensive to conduct properly devised probability surveys, on accounts of recruiting interviewers, pursuing selected respondents for a continued period with declining response rates and other reasons. Such reasons, alongside the influx of big data in the age of advanced computing, contribute to the newly gained popularity of nonprobability surveys; although they have {existed} in the sample survey literature for a very long time. When quick responses can be obtained through web surveys or any individual can be sampled just maintaining some overall quotas (as in quota sampling), there is increased inclination towards using these more convenient options. 
Reflecting this growing interest, recent special issues of journals -- Survey Methodology and the Calcutta Statistical Association Bulletin, have been dedicated to statistical research in nonprobability sampling.

Care must be taken while producing estimates for the targeted finite population characteristics based on such nonprobability surveys because of their obvious selection bias issue. Researchers have devised various statistical methodologies for improving representativeness of nonprobability surveys by integrating them with probability surveys (termed as `reference surveys'), which usually do not have the response variables of interest but have useful auxiliary variables. For an overview, we refer to \cite{rao2021}, \cite{beaumont2020}, among others. Specifically when common auxiliary variables are available for both types of surveys, some commonly used methods include matching and mass imputation methods (\cite{rivers2007}, \cite*{kim2021}), {inverse propensity score weighting (IPW)} methods (\cite{leevalliant2009}, \cite{valliant2011}, \cite*{Wang}), doubly robust methods (\cite*{Chen}) and others. \cite{Wu} provides an excellent review of these model-based approaches and related assumptions. More recently, \cite*{beaumont2024} have investigated variable selection techniques in conjunction with propensity score weighting and post stratification of estimated weights. In contrast to these methods that require combining both probability and nonprobability surveys, \cite{pfeffermann2015} and \cite{kimcsa} rely solely on nonprobability surveys for inference.

It is well-established in the literature that applying data integration methods make nonprobability surveys more representative of the finite population, but such methods are developed under the assumption that the survey responses are accurate. \cite{KML} point out that {a significant amount of measurement error is present in online commercial nonprobability surveys} (referred to as opt-in, in short) and bias owing to reasons {other than sampling error} should not be overlooked. This research is motivated by the 2021 Benchmarking Study (\cite{PEW}), designed by the Pew Research Center (referred to as Pew, in short) to determine accuracy of online surveys on general population estimates for all adults in the United States (U.S.) as well as key demographic subgroups such as age, race, gender, education, etc. The benchmarks are obtained from large government surveys such as the American Community Survey (ACS), the Current Population Survey (CPS), among others. \citet{KML} observe that even after calibration, responses from certain subgroups in opt-in surveys remained highly biased—unlike in probability surveys, where the corresponding estimates are closely aligned with benchmark values. The reason behind this is attributed to `bogus responding', {e.g. replying `Yes’ regardless of the question}. These responses are basically insincere / fraudulent survey answers, causing high measurement error in opt-in surveys. For details on the bogus responding literature, we refer the reader to \cite{bogus}. 

In \cite{KML}, the authors use benchmarking to identify subgroups where bogus responding is concentrated in opt-in surveys, but they do not propose a statistical solution on how to fix this problem. A quick fix by removing the bogus respondents altogether may make matters worse. The authors highlight the need for statisticians working with opt-in samples to address both {issues of} bogus responding and representativeness – not just the latter. 
Our primary objective is to integrate two important areas of research in nonprobability sampling: improving representativeness and addressing measurement differences. In our framework, both the probability and nonprobability surveys contain the response variables of interest, as well as a set of common and informative auxiliary variables. The availability of response variables in both surveys is essential for correcting bias arising from measurement differences. In {Section \ref{sec:res_qs}}, we describe our main research questions and our proposed solutions to answer these questions. We provide a detailed discussion about assumptions in {Section \ref{sec_cond}}. {For simplification hereon, we refer to probability surveys as $ps$ and nonprobability surveys (specifically opt-in surveys) as $nps$.} 

\subsection{Research questions}
\label{sec:res_qs}

\textit{\begin{itemize}
    \item [\hypertarget{Qs.1}{Qs.1}] Does improving representativeness alone improve estimates from $nps$?
\end{itemize}}
To answer \hyperlink{Qs.1}{Qs.1}, we apply the IPW method of \cite{Chen}{, assuming that IPW adequately adjusts the selection bias from $nps$. We provide {a review of the IPW method} in Section \ref{sec_rep}. For more details on the assumptions (such as missing at random (MAR)), we refer the reader to \cite{Wu}. 
It is worth pointing out that} \cite{KML} only used benchmarking techniques on the case study and model-based methods such as IPW were not used to improve representativeness in their $nps$. It is also important to note that competing weighting methods, as discussed in earlier paragraphs in the introduction, may perform differently. We adopt IPW, which is one of the widely used methods, and do not delve further into comparisons with other methods, as it does not fall within the main focus of this article.
\textit{\begin{itemize}
    \item [\hypertarget{Qs.2}{Qs.2}] How to correct for measurement differences in $nps$?
\end{itemize}}
For \hyperlink{Qs.2}{Qs.2}, we devise a methodology to correct for the measurement differences in $nps$ due to bogus responding. The benchmarking study in \cite{KML} reveals that when compared to large surveys (CPS, ACS, etc.), $ps$ are not subject to high measurement errors, but $nps$ are. Our aim in this article, is to bring the error levels in $nps$ at par with $ps$. Hence, following the footsteps of \cite{KML}, {we consider $ps$
to be the benchmark, since it is assumed that measurement error is smallest in this data set. In this way the proposed method corrects for relative measurement bias between $ps$ and $nps$ (not absolute bias).}
While we recognize {that this assumption may not hold perfectly in practice}, our focus lies in enhancing the quality of $nps$ by leveraging $ps$, rather than refining $ps$ themselves. Consequently, our proposed method to address  \hyperlink{Qs.2}{Qs.2} is centered on correcting `measurement differences' between the two types of surveys.
\textit{\begin{itemize}
    \item [\hypertarget{Qs.3}{Qs.3}] How to combine estimates from $ps$ and $nps$ 
    for producing estimates with lower bias? 
\end{itemize}}
We empirically observe that even after correcting for measurement differences, the finite population mean estimates from $nps$ can be further improved through carefully designed composite estimators. Such estimators have been studied in small area estimation (\cite{schaible1978}, \cite{Rao2015}), where the authors consider a weighted combination of a direct and a synthetic estimator. \cite{Elliot} also propose a weighted estimator based on an $nps$ (web-based convenience sample) with a $ps$ to achieve a smaller Mean Squared Error (MSE) than estimators based on $ps$ alone. More recently, \cite{KimTam} propose a bias-corrected data integration estimator, leveraging big data (which can also be thought of as an $nps$) and one $ps$, where one of them is subject to measurement error.

To address \hyperlink{Qs.3}{Qs.3}, we propose a new composite estimator, which is a weighted combination of unbiased estimates from both $ps$ and $nps$, aiming to reduce both selection bias and measurement differences in $nps$. In this paper, we show analytically that the MSE of the proposed estimator is lower than that of \cite{Elliot}. Additionally, 
we incorporate predictive modeling using advanced machine learning (ML) models (Random Forest, Gradient Boosting) to adjust for bias in IPW-based estimates. ML models are becoming increasing popular in sample surveys. Recent survey sampling literature, \cite*{toth2011}, \cite*{Breidt2017}, \cite*{kern2019}, \cite*{dagdoug2021}, among others, use modern prediction techniques such as Regression Trees, Random Forest, etc., for model assisted survey estimation.

The estimator proposed in \cite{KimTam} accounts for both selection bias and measurement errors in big data without assuming MAR in their missing data approach. However, the method requires identification of overlapping units and uses calibration weighting, treating the big data as an incomplete sampling frame for the finite population. In contrast, our method is simpler (not needing unit-level comparison)  but relies on the MAR assumption for IPW-based bias correction. Structurally, our estimator is more aligned with that of \cite{Elliot}, but differs in the sense that it minimizes MSE over a different estimator class. Moreover, \cite{Elliot} attribute the bias mainly from the point of view of sampling error (i.e., selection bias) without considering measurement error that is also closely aligned with our framework. 
\textit{\begin{itemize}    
    \item [\hypertarget{Qs.4}{Qs.4}] When do composite estimators outperform estimators from $ps$ alone? 
\end{itemize}}
From our data analysis, we observe that the proposed composite estimator does not uniformly outperform estimates based solely on the $ps$, when the $ps$ has a sufficiently large sample size. Hence, to answer \hyperlink{Qs.4}{Qs.4}, we draw smaller samples from an available $ps$ and combine with an $nps$, to demonstrate a scenario where proposed composite estimator outperforms estimator created only using $ps$.

\subsection{Conditions and assumptions}
\label{sec_cond}
Having outlined our research questions and proposed solutions, we now proceed to a detailed discussion on the key assumptions underlying our methodology. The proposed approach primarily relies on two central assumptions, \hyperlink{A1}{(A1)} and \hyperlink{A2}{(A2)}:
\begin{itemize}
    \item [\hypertarget{A1}{(A1)}] IPW primarily removes selection bias from nonprobability samples.
\end{itemize}        
This assumption forms the basis of our approach to \hyperlink{Qs.1}{Qs.1}, which focuses on improving the representativeness of $nps$. IPW method has been extensively used in literature for this purpose. However, this method assumes a MAR selection mechanism (\cite{rubin1976}).
If MAR assumption doesn't hold, then estimating propensity scores is very challenging (\cite{little2019}). 
Moreover, IPW requires strong auxiliary information to correct for selection bias. In the absence of good covariates, residual selection bias may persist in $nps$ even after applying IPW. Specifically, the differences between $ps$ and $nps$ after correcting for selection bias will be a combination of residual selection bias in $nps$, measurement errors in $nps$ and $ps$, and sampling error. However, we do not further explore such a situation for this analysis, as our main purpose is to find new techniques to integrate the two areas of bias (selection and measurement) in $nps$. 
\begin{itemize}
    \item [\hypertarget{A2}{(A2)}] {Probability samples have negligible measurement error.}
\end{itemize}  
This assumption is used in correcting for measurement differences in $nps$. The presence of measurement errors in $ps$ is very likely, as any survey, however carefully conducted, is subject to measurement error. Extensive literature has addressed measurement errors in probability surveys, starting from the work of \cite{mahalanobis1946recent}, in the context of agricultural surveys and crop estimation in India. {\cite*{Biemer} in his book chapter refers} to the classic work by \cite*{hansen1961measurement} for estimation of measurement error variance. \cite*{groves} discuss measurement errors in $ps$ within the `Total survey error' framework. 

Several articles discuss that mode effects causing measurement errors in probability surveys can be reduced by good questionnaire design or adjusted by weighting / matching methods, after the survey has been conducted. Specifically, \cite*{jackle2010} discuss methods for assessing mode effects using proportional odds models, structural equation modeling, etc. \cite*{suzer2012} use multiple imputation estimation to analytically isolate mode choice and measurement effects. \cite*{SCHOUTEN2013} propose a new experimental design that allows for a decomposition of relative mode effects. For more details see \cite*{van2010}, \cite*{buelens2015}, \cite*{Brakel2} and the references cited therein.
In essence, evaluating and correcting measurement errors in $ps$ is a vast research area, which is not focus of this article. In this article, we are mainly concerned with improving $nps$ by leveraging available $ps$. Hence, we do not pursue above-mentioned methods further and leave this topic for future research. 

The rest of the paper is organized as follows. In Section \ref{sec_rep}, we provide a review of the IPW estimator from \cite{Chen}, used for improving representativeness in $nps$. In Section \ref{sec_comp}, we discuss our proposed method of treating selection bias and measurement {differences} together through composite estimators, devised by integrating $ps$ and $nps$, when both surveys have the response variable/s of interest. In Section \ref{sec_data}, we describe the motivating case study from Pew and provide necessary details on the surveys, including data collection method, sample size, benchmarks, weighting, etc. In Section \ref{sec_analysis}, we provide the results of our data analysis, sequentially providing answers to the four research questions raised in the earlier paragraphs. Finally, we end with concluding remarks in Section \ref{sec_conclusion}.

\section{Improving representativeness}
\label{sec_rep}

The main objective of this section is to address the representativeness issue of $nps$ and improve said selection bias. To this end, we use the IPW method described in \cite{Chen}. IPW is computationally advantageous as it produces one set of estimated weights {that} can be used for any response variable. This is similar to the design-based approach of the Horvitz-Thompson estimator, the difference here being that the selection probabilities are estimated instead of being known from the sample design. To use this method, information on common auxiliary variables (from both $ps$ and $nps$) and survey weights (from $ps$) are required. For completeness, we provide a review of the IPW method described in \cite{Chen} in the following paragraph. 

Let $S_A$ be an $nps$ of size $n_A$ and $S_B$ be a $ps$ of size $n_B$ from a finite population of size $N$. $d_i^B$ is the known design weight of $i^{th}$ unit from $ps, \; i = 1, \ldots , n_B$. The selection indicator for $S_A$, given by $R_i$, is defined as follows:
\begin{align*}
        R_i =
        \begin{cases} 
        1 & \mbox{if } i \in S_A, \\
        0   &  \mbox{if } i \notin S_A,
    \end{cases} 
    \quad i=1,\ldots,N.
    \end{align*}
The propensity score for the $i^{th}$ unit is defined as $\pi_i^A = P_q(R_i = 1 | x_i), \; i = 1,\ldots, N$, where $q$ is the model for the selection mechanism for $S_A$ and $x$ is vector of covariates. Following assumptions from \cite{Chen}, we have $\pi^A_i > 0$, $R_i$ and response $y_i$ are independent given $x_i$, $R_i$ and $R_j$ are independent given $x_i$ and $x_j$, for all $ i \ne j$.
The propensity score takes the following form for {the} logistic model: 
\begin{align*}
\pi_i^A = \pi(x_i, \theta_0) = \frac{\exp(x_i'\theta_0)}{1 + \exp(x_i'\theta_0)}, \; i = 1,\ldots, N,
\end{align*}
where $\theta_0$ is the true value of the unknown model parameters. Maximum Likelihood Estimate (MLE) of the propensity score is
$\hat{\pi}^A_i = \pi(x_i,\hat{\theta})$, where $\hat{\theta}$ maximizes the following pseudo-log-likelihood function:
\begin{align*}
    l^*(\theta) \equiv \sum_{i \in S_A} x_i'\theta - \sum_{i \in S_B} d_i^B \log \lbrace 1 +  \exp(x_i'\theta) \rbrace.   
\end{align*}
Solution of the above equation is obtained using the Newton-Raphson iterative procedure. Finally, for a response variable $y$, the IPW estimator for the population mean $\mu_Y = N^{-1} \sum_{i=1}^{N} y_i$ is given by 
\begin{align} \label{eq:IPW}
  \hat{\mu}_{CLW} \equiv (\hat{N}^{A})^{-1} \sum_{i \in S_A} (y_i/\hat{\pi}^A_i),  
\end{align}
where $\hat{N}^A = \sum_{i \in S_A} (\hat{\pi}^A_i)^{-1}$. For detailed expression of the variance of $\hat{\mu}_{CLW}$, we refer to \cite{Chen}. 

For $ps$, we use the survey weighted estimator for response variable $y$, denoted by
\begin{align} \label{eq:ps}
  \hat{\mu}_{{cal}} \equiv (\hat{N}^{B})^{-1} \sum_{i \in S_B} d_i^B y_i, 
\end{align}
where $\hat{N}^B = \sum_{i \in S_B} d_i^B$. 
The uncertainty measure of this estimator, 
can be derived from the design information. The estimator defined in \eqref{eq:ps} can also be used for constructing a weighted estimator from $nps$, using calibrated weights instead of design weights. Hence, to avoid confusion, we use the abbreviation `cal' in the estimator defined in \eqref{eq:ps}, as we will be using the same estimator for $ps$ (with design weights) and $nps$ (with calibrated weights by Pew) in our data analysis.

\section{Composite estimators}  
\label{sec_comp}

In this section, we introduce a method to combine estimates from $ps$ and $nps$, given that the response variable of interest is present in both the surveys. 
To formulate this composite estimator, we adopt the model described in \cite{Elliot}, where bias in $nps$ is assumed to be known. Mathematically, let $\bar{y}_1$ be a survey weighted estimator of population mean from $ps$ ($\hat{\mu}_{{cal}}$ in our case), $\bar{y}_2$ an estimator from $nps$ ($\hat{\mu}_{CLW}$ in our case). Let $\mu$ be the true mean of the $ps$ estimator, whereas $nps$ estimator has an additional bias term $\epsilon$. In this model, $\epsilon$ is assumed to be known and $\mu$ is the unknown parameter of interest. $v_1$ and $v_2$ are known variances of $\hat{\mu}_{{cal}}$ and $\hat{\mu}_{CLW}$ respectively. The correlation between estimates from $ps$ and $nps$ is assumed to be $0$, as the $ps$ and $nps$ are independently drawn. The sampling model of \cite{Elliot} can be written as:
\begin{align} \label{eq:model}
    \begin{pmatrix}
        \bar{y}_1 \\ \bar{y}_2
    \end{pmatrix} 
    \sim 
    \begin{pmatrix} \begin{pmatrix} \mu \\ \mu + \epsilon  \end{pmatrix},  \begin{pmatrix} v_1 & 0 \\ 0 & v_2 \end{pmatrix} \end{pmatrix}. 
    \end{align}
In practice, a plug-in estimator of $\epsilon$, obtained from historical data or subject matter knowledge, can be used. \cite{Elliot} propose to estimate bias in practice as the difference between the mean estimates of $ps$ and $nps$.
$v_1$ and $v_2$ can be obtained from the estimates of variance of a design-based estimator (such as our $\hat{\mu}_{{cal}}$) and the estimate of variance of $nps$ (such as $\hat{\mu}_{CLW}$ derived in \cite{Chen}), respectively.
    
In \cite{Elliot}, the authors propose a composite estimator of the form:
\begin{align} \label{eq:EV}
    \hat{\mu}_{EV} & \equiv \frac{(\epsilon^2 + v_2)\bar{y}_1 + v_1 \bar{y}_2}{\epsilon^2 + v_1 + v_2},
\end{align}
which is a biased estimator, with remaining bias given by 
\begin{align}\label{eq:bias}
    {b_{EV}} \equiv E(\hat{\mu}_{EV}) - \mu 
    %&= \frac{(\epsilon^2 + v_2)\mu + v_1(\mu + \epsilon)}{\epsilon^2 + v_1 + v_2} - \mu \\ 
    &= \frac{\epsilon v_1}{\epsilon^2 + v_1 + v_2},
\end{align}
and variance given by 
\begin{align} \label{eq:var}
    \mbox{Var}(\hat{\mu}_{EV}) = \frac{v_1\lbrace (\epsilon^2 + v_2)^2 + v_1 v_2 \rbrace}{(\epsilon^2 + v_1 + v_2)^2}.
\end{align}
In the above equations, we use the subscript `EV' after the authors' names, for simplified notation.
Therefore, from \eqref{eq:bias} and \eqref{eq:var}, MSE of $\hat{\mu}_{EV}$ is given by
\begin{align} \label{eq:MSE_EV}
    \MSE(\hat \mu_{EV}) \equiv b_{EV}^2 + \mbox{Var}(\hat{\mu}_{EV}) = \frac{v_1(\epsilon^2 + v_2)}{\epsilon^2 + v_1 + v_2}.
\end{align}
$\hat{\mu}_{EV}$, as defined in \eqref{eq:EV}, is the most efficient estimator in the sense of minimizing MSE in a class of biased estimators.

We propose an unbiased composite estimator, which is a weighted combination of mean estimates from $ps$, given by $\hat{\mu}_{{cal}}$, and `bias-corrected' estimates from $nps$, given by $\hat{\mu}_{bc;CLW} \equiv \hat{\mu}_{CLW} - \epsilon$, {where subscript `bc' stands for bias-corrected.} This composite estimator, {denoted by $\hat{\mu}_{comb}$ with subscript standing for `combined',} is defined as follows
{
\begin{align}
    \hat{\mu}_{comb} (w) \notag
    &\equiv w \hat{\mu}_{cal} + (1-w)\hat{\mu}_{bc;CLW}\\
    &= w \bar{y}_1  + (1-w)(\bar{y}_2 - \epsilon), \label{eq:mu-comb}
\end{align}    
}
where the weight {$w\in (0,1)$} is determined by minimizing the MSE of {$\hat{\mu}_{comb}(w)$} given by
\begin{align} \label{eq:MSE_Comb1}
    {\MSE(\hat \mu_{comb}(w))} &\equiv E[w \bar{y}_1  + (1-w)(\bar{y}_2 - \epsilon) - \mu]^2 \notag \\
    &= w^2 v_1 + (1-w)^2 v_2.
\end{align}

Taking derivative of \eqref{eq:MSE_Comb1} with respect to $w$ and equating to $0$ we get ${ w_1^*}=v_2/(v_1 + v_2)$.
Finally, we obtain the expression:
\begin{align} \label{eq:comb}
    \hat{\mu}_{comb}
    &\equiv \Bigg( \frac{v_2}{v_1 + v_2} \Bigg) \bar{y}_1 + \Bigg ( \frac{v_1}{v_1 + v_2} \Bigg) (\bar{y}_2 - \epsilon).
\end{align}
The proposed estimator in \eqref{eq:comb} is unbiased {for $\mu$}, as it combines two unbiased estimators: $\hat{\mu}_{{cal}}$ and $\hat{\mu}_{bc;CLW}$. MSE of $\hat{\mu}_{comb}$ is given by
\begin{align}\label{eq:MSE_Comb}
\MSE(\hat \mu_{comb}) = \frac{v_1 v_2}{v_1 + v_2}.
\end{align}
$\hat{\mu}_{EV}$ and $\hat{\mu}_{comb}$ both minimize MSE, but in disjoint classes of estimators. To compare $\hat{\mu}_{EV}$ and $\hat{\mu}_{comb}$, we compare the respective MSE's given in \eqref{eq:MSE_EV} and \eqref{eq:MSE_Comb}.
Note that {since $\epsilon^2 > 0$}, we have the following
\begin{align*}
    v_2(\epsilon^2 + v_1 + v_2) = (\epsilon^2 + v_2)v_2 + v_1 v_2 <  (\epsilon^2 + v_2)v_2 + v_1 (\epsilon^2 + v_2) = (\epsilon^2 + v_2) (v_1 + v_2).
\end{align*}
From above, since 
\begin{align}
    & \frac{v_1 v_2}{v_1 + v_2} < \frac{v_1(\epsilon^2 + v_2)}{\epsilon^2 + v_1 + v_2}, \label{eq:mse-comparison}
\end{align}
we have $\MSE(\hat\mu_{comb}) < \MSE(\hat\mu_{EV})$.

We note that in the previous discussion, the bias term $\epsilon$ needs to be known possibly from alternative sources. Assume that $\hat \epsilon$, an estimator of $\epsilon$, is constructed from the auxiliary survey that is independent of $\bar y_2$. Moreover, let $E(\hat \epsilon)= \epsilon$, $\text{Var}(\hat \epsilon) = v_b$. Under these assumptions, the combined plug-in estimator can be written as 
\begin{align}
    \tilde \mu_{comb} (w) \equiv w \bar y_1 + (1- w) (\bar y_2 - \hat \epsilon) \label{eq:new-mu-comb},
\end{align}
where the weight $w \in (0,1)$.
Using the unbiasedness of $\hat \epsilon$ it follows that $E(\tilde \mu_{comb}(w)) =\mu$, i.e., $\tilde \mu_{comb}(w)$ is an unbiased estimator of $\mu$ for all values of $w \in (0,1)$.
The corresponding MSE of $\tilde \mu_{comb}(w)$
\begin{align}
    \MSE( \tilde \mu _{comb}(w)) = w^2 v_1 + (1-w)^2 (v_2 + v_b). \label{eq:mse-new-mu-comb}
\end{align}
Minimizing the MSE in \eqref{eq:mse-new-mu-comb} with respect to $w$ yields the optimal weight $ w_2^* = (v_2 + v_b)/(v_1+v_2+v_b)$. Therefore, substituting this optimal weight into \eqref{eq:new-mu-comb} yields the plug-in estimator:
\begin{align}
    \tilde \mu_{comb} \equiv \left( \frac{v_2+v_b}{v_1+v_2+v_b} \right) \bar y_1 + \left( \frac{v_1}{v_1+v_2+v_b} \right) (\bar y_2 - \hat \epsilon) \label{eq:new-mu-comb-2}
\end{align}
and the MSE of $\tilde \mu_{comb}$ is given by $\MSE(\tilde \mu_{comb}) = v_1(v_2+v_b)/(v_1+v_2+v_b)$. It is worth noting that $\hat \mu_{comb}$ is a special case of $\tilde \mu_{comb}$ with $v_b =0$, corresponding to the degenerate case where the bias estimator satisfies $\hat \epsilon = \epsilon$ with probability 1.
When compared with the estimator proposed in \citet{Elliot}, similar algebraic manipulation as in \eqref{eq:mse-comparison} shows that $\MSE(\tilde \mu_{comb}) \leq \MSE(\hat \mu_{EV})$ provided $v_b \leq \epsilon^2$. This suggests that the plug-in estimator performs better than \cite{Elliot} when the variance of the bias estimator is relatively small compared to the square of the bias. A specific construction of the bias estimator $\hat \epsilon$ is presented in the real data application in Section \ref{sec_mebc}.

%%%%%%%%%%%%%%%%%%%%%%%%%%%%%%%%%%%%%%%%%%%%%%%%%%%%%%%%%%%%%%%
\section{Data description}
\label{sec_data}

For this analysis, we use six surveys sourced from Pew's 2021 Benchmarking Study. The data and detailed report are available on {the Pew website \href{https://www.pewresearch.org/methods/2023/09/07/comparing-two-types-of-online-survey-samples/}{(weblink)}}. The surveys in this study were administered between June $14$ and July $21, 2021$. They included interviews with a total of $29,937$ U.S. adults and approximately $5,000$ in each sample. Three of the surveys were sourced from different probability-based online panels, one of which was Pew’s American Trends Panel (ATP). The remaining three surveys came from three different online opt-in sample providers. These surveys were conducted through online platforms. 

Table \ref{tab: KML_info}, sourced from \cite{KML}, provides detailed information on the sample sizes and field dates. A common questionnaire was administered in English or Spanish for all six surveys. Information on demographics such as age, race, gender, education, marital status were obtained along with other answers on insurance coverage, smoking status, blood pressure level, U.S. citizenship, number of kids/adults in households, etc. These six survey datasets can be categorized into two types: nonprobability surveys ($nps$) and probability surveys ($ps$), as follows:

\textbf {i. Nonprobability surveys ($nps$):} Three of the six surveys are commercial opt-in surveys, denoted as Opt-in 1,2,3. These were obtained from vendors who used quota sampling. Pew provided vendors with quota targets for age $\times$ gender, race $\times$ Hispanic ethnicity, education from the 2019 ACS. Ipsos was the data collection firm in this case. Since these are online nonprobability panels, we do not have known selection probabilities assigned to the individuals.

\textbf {ii. Probability surveys ($ps$):} The other three are probability surveys. Although the panels are online, panelists are recruited offline, {e.g. using probability sampling}. To be specific, they are recruited using address-based sampling (ABS) from the U.S. Postal Service Computerized Delivery Sequence File. Hence, following the nomenclature of \cite{KML}, we refer to these surveys as ABS 1,2,3. The study-specific response rates of ABS 1,2,3 were $61\%$, $90\%$, and $71\%$, respectively. 

\begin{table}[ht]
        \caption{Sample Sizes and Field Dates of six Pew surveys - sourced from \cite{KML}.}
        \label{tab: KML_info}
        \centering
        \begin{tabular}{ccccc}
        % \hline
        \toprule
        Sample type & Sample Name & ID & Sample size & Field dates\\
        % \hline
        % \hline
        \midrule
         & ABS panel 1 & $P_1$ & 5,027 & June 14 - 28, 2021\\
        Probability & ABS panel 2 & $P_2$ & 5,147 & June 14 - 27, 2021\\
         & ABS panel 3 & $P_3$ & 4,965 & June 29 - July 21, 2021\\
         \hline
         & Opt-in panel 1 & $O_1$ & 4,912 & June 15 - 25, 2021\\
        Nonprobability & Opt-in panel 2 & $O_2$ & 4,931 & June 11 – 27, 2021\\
         & Opt-in panel 3 & $O_3$ & 4,955 & June 11 – 26, 2021 \\ 
         % \hline
         \bottomrule
    \end{tabular} 
    \end{table}

\textbf {Weighting}: \cite{KML} used a standard weighting approach of calibration for all six surveys. For the nonprobability surveys since there was no design weight a starting weight of $1$ was assigned to every individual. For probability surveys, panel base weights were adjusted for differential probabilities of selection and then used in calibration. Multiple population targets (from ACS, CPS and Pew's National Public Opinion Reference Survey) such as age $\times$ sex, race/ethnicity $\times$ education, etc., were used to calibrate the starting weights for each sample to a common set of population control totals.

\textbf {Benchmark}: Some of the survey questions were considered for benchmarking analysis where the benchmarks were obtained from sources such as National Health and Nutrition Examination Survey (NHANES), National Health Interview Survey (NHIS), ACS, CPS, Pew Survey on COVID data tracker and Election projection data. These values are provided at national level as well as by categories of demographic variables such as age (3 categories viz. $18-29$, $30-64$, $65+$ years), race (3 categories viz. White non-Hispanic, Black non-Hispanic and Hispanic) and education (3 categories viz. Higher Secondary (HS) or less, Some college and College graduate). For the analysis of measurement error, these benchmark values are compared with survey estimates (calculated from the six surveys) at these subgroup levels of demographic variables or overall survey level, for response variables of interest. We elaborate this in the following Evaluation Section.

\subsection{Evaluation}

To evaluate the estimators in terms of both representativeness bias and measurement differences, we use the Mean Squared Deviation/Difference (MSD) as the evaluation criterion -- lower MSD implying better performance of an estimator. We define this error in terms of deviation from benchmark values as follows. 
Let $m$ be the number of benchmark variables for a survey ($ps$ or $nps$) and $k$ be the number of subgroups based on factors of interest (viz. demographic variables such as age, race etc). If $\bar{Y}_{jc}$ is the value for question $j$ and subgroup $c$ from benchmark source (considered as true value) and $\hat{\bar{Y}}_{jc}$ is the respective survey estimate (where $j = 1, \cdots, m; \; c = 1, \cdots, k$), then {MSD} for the estimator $\hat{\bar{Y}}$ is defined as
    \begin{align} \label{eq:MSE}
        \MSD(\hat{\bar{Y}}) \equiv m^{-1}\sum_{j=1}^{m} \left [{k}^{-1}\sum_{c=1}^{k}\left(\hat{\bar{Y}}_{jc} - \bar{Y}_{jc}\right)^2  \right ].
    \end{align}
As an example we mention that, to evaluate the impact of selection bias improvement we compute $\hat{\mu}_{CLW}$ from $nps$ and calculate {$\MSD(\hat{\mu}_{CLW})$}. We compare {$\MSD(\hat{\mu}_{CLW})$} with {$\MSD(\hat{\mu}_{cal})$} from $ps$. Evaluations can be done at an overall survey level like in \eqref{eq:MSE} or at a subgroup level of any factor of interest or at question level or both, removing the averaging factor accordingly. In this data analysis, we only consider binary questions and do not consider the level of a question as an additional component in MSD. 

\section{Data analysis}
\label{sec_analysis}

Before proceeding into our empirical investigation we take into account the key findings in \cite{KML}. The authors use the measure Mean Absolute Error (MAE) and observe that MAE values are higher in opt-in surveys as compared to ABS. At the subgroup level, MAE values show a distinct pattern in opt-in surveys that is absent in ABS surveys. For example, among $18-29$ year-old's and Hispanic adults, MAE values are higher in opt-in surveys, likely due to `bogus-responding'. Bogus responding is more prominent in questions regarding government benefits such as social security, food stamps, unemployment compensation or workers’ compensation. Large shares of $18-29$ year-old's and Hispanic adults report having received at least three of four such benefits, which is very rare in the true population. According to \cite{KML}, for opt-in surveys, this ranges between $6\%$ to $11\%$, while the true population incidence is $0.1\%$. These insightful findings motivate us to use advanced methods for weighting in opt-in samples in conjunction with bias-correction due to measurement differences. In the following sections, we describe our data analysis methods. All of our codes are available at this website \href{https://github.com/asen123/nonprob}{(weblink).}

\subsection{IPW estimators from opt-in samples}
\label{sec:IPW_resutls}
We use the method discussed in Section \ref{sec_rep} leveraging \texttt{R} package \texttt{nonprobsvy} to calculate $\hat{\mu}_{CLW}$ for 3 opt-in samples $O_1, O_2, O_3$, with $P_1, P_2, P_3$ respectively as reference. We emphasize that there is no inherent pairing between the surveys $O_j$ and $P_j$; that is, any combination of $O_j$ and $P_k$, for $j, k = 1, 2, 3$, can be used to improve representativeness. In our analysis, we use the matched indices $O_j$ and $P_j$ for $j = 1, 2, 3$, but we expect similar results with other combinations. This assignment also ensures independence across the pairs, which is critical for estimating $\hat \epsilon$ in the composite estimator \eqref{eq:new-mu-comb-2} using external data sources independently.
In the propensity score logistic model we consider age, gender, race, education and region as auxiliary variables. For opt-in samples, we compare two estimators {in terms of MSD} -- one is survey weighted mean using calibrated weights from Pew ($\hat{\mu}_{cal}$) and the other is IPW estimator ($\hat{\mu}_{CLW}$). For ABS samples, only the first one is applicable ($\hat{\mu}_{cal}$). 
A list of the estimators that we compare in this data analysis, with detailed formula, is provided in Table \ref{est_list}. For these estimators, {MSD} values are calculated using \eqref{eq:MSE} with $m=12$ benchmark questions having only `Yes'/`No' answer. Details of these questions are provided in Table \ref{qs_det}. 

\begin{table}[!htb]
\centering
\caption{List of finite population mean estimators, with equation number for detailed formula and types of surveys. In column 5, `or' implies that an estimator is defined for either $ps$ or $nps$ and `$\&$' implies that an estimator is defined using both $ps$ and $nps$.}
\label{est_list}
\begin{minipage}{\textwidth}
\begin{tabular}{lllcc}
  % \hline
  \toprule
 No. & Estimator & Details of the estimation method & Formula  & Survey type\\ 
  % \hline
  \midrule
  $1$ & $\hat{\mu}_{{cal}}$ & \begin{tabular}[c]{@{}l@{}} Weighted mean using design or \\ calibrated weights \end{tabular} &
  \eqref{eq:ps} & $ps$ or $nps$\\\\
  $2$ & $\hat{\mu}_{CLW}$ & IPW estimator of \cite{Chen} & \eqref{eq:IPW} & $nps$\\\\
  $3$ & $\hat{\mu}_{bc;CLW}$ & \begin{tabular}[c]{@{}l@{}} Measurement difference bias-corrected \\ version  of $\hat{\mu}_{CLW}$, where bias is \\calculated from \eqref{eq:eps} \end{tabular}  & \eqref{eq:bc} & $nps$\\\\
  %$4$ & $\hat{\mu}_{cal}$ & \begin{tabular}[c]{@{}l@{}} Weighted mean using calibrated \\ weights from Pew \end{tabular} & \eqref{eq:ps} & $ps$ or $nps$\\\\
  $4$ & $\hat{\mu}_{EV}$ & \begin{tabular}[c]{@{}l@{}} Composite estimator of \\ \cite{Elliot}, \\
  where bias is calculated from \eqref{eq:eps} \end{tabular} & \eqref{eq:EV} & $ps$ \& $nps$\\\\ 
  $5$ & {$\tilde{\mu}_{comb}$} & \begin{tabular}[c]{@{}l@{}} Proposed composite estimator \\
  combining {$\hat \mu_{cal}$ and $\hat{\mu}_{bc;CLW}$}, \\
  where bias is calculated from \eqref{eq:eps} \end{tabular} & \eqref{eq:new-mu-comb-2} & $ps$ \& $nps$\\\\ 
  $6$ & $\hat{\mu}_{{ML};CLW}$ & \begin{tabular}[c]{@{}l@{}}IPW estimator of \cite{Chen} \\
  using predicted responses\\ {(where `ML' stands for model)} \end{tabular}  & \eqref{eq:IPW2} & $ps$ \& $nps$\\\\ 
  $7$ & $\hat{\mu}_{{ML};comb}$ & \begin{tabular}[c]{@{}l@{}} Proposed composite estimator \\ combining $\hat{\mu}_{{cal}}$ and $\hat{\mu}_{{ML};CLW}$ \end{tabular} & \eqref{eq:comb_m} & $ps$ \& $nps$\\\\
  $8$ & $\hat{\mu}_{{ML};EV}$ & \begin{tabular}[c]{@{}l@{}} Version of $\hat{\mu}_{EV}$, where bias is \\ difference of $\hat{\mu}_{{cal}}$ and $\hat{\mu}_{CLW}$ \end{tabular} & \eqref{eq:EVm} & $ps$ \& $nps$\\
   \bottomrule
\end{tabular}
\end{minipage}
\end{table}

\begin{table}
\small{
    \caption{Details of $12$ questions (with only `Yes'/{`No'} answers) used as benchmark. Sourced from Questionnaire on the Pew Research Centre website \href{https://www.pewresearch.org/methods/2023/09/07/comparing-two-types-of-online-survey-samples/}{(weblink).}} \label{qs_det}
\centering
\begin{tabular}{ll}
% \hline
\toprule
    Question No. & Question Wording\\
    \midrule
    1. Insurance & \parbox{10cm}{Are you currently covered by any form of health insurance \\ or health plan?}\\\\
    2. Blood pressure & \parbox{10cm}{Have you ever been told by a doctor or other health \\ professional that you had hypertension, also called high \\ blood pressure?} \\\\
    3. Parent & Are you the parent or guardian of any children under age 18? \\
    4. Food allergy & Do you have any food allergies? \\
    5. Job last year & Did you work at a job or business at any time during 2020? \\\\
    6. Retirement account & \parbox{10cm}{At any time during 2020 did you have any retirement \\ accounts such as a 401(k), 403(b), IRA, or other account \\ designed specifically for retirement savings?} \\ \\
    7. Unemployment compensation & \parbox{10cm}{At any time during 2020, did you receive any State or \\ Federal unemployment compensation?} \\ \\
    8. Workers' compensation & \parbox{10cm}{During 2020 did you receive any Worker’s Compensation \\ payments or other payments as a result of a job-related \\ injury or illness?} \\\\
    9. Food stamps & \parbox{10cm}{At any time during 2020, did you or anyone in your \\ household receive benefits from SNAP (the \\ Supplemental Nutritional Assistance  Program) \\ or the Food Stamp program, or use a SNAP or \\ food stamp benefit card?} \\\\
    10. Social Security & \parbox{10cm}{During 2020 did you receive any Social Security payments \\ from the U.S. Government?} \\\\
    11. Union membership & \parbox{10cm}{Are you a member of a labor union or of an employee \\ association similar to a union?} \\\\
    12. U.S. citizenship & Are you a citizen of the United States? \\        
    \bottomrule
\end{tabular}
}
\end{table}

% \begin{table}[ht]
% \centering
% \caption{\AB{MSD } values (in \%) of population mean estimators, calculated using $12$ benchmark questions from $6$ surveys. Refer to Table \ref{est_list} for more details on the estimators.}
% \label{tab:MSE_all}
% \begin{tabular}{lrrrrrr}
%   \toprule
%  Estimator & $P_1$ & $P_2$ & $P_3$ & $O_1$ & $O_2$ & $O_3$ \\ 
%   % \hline
%   \midrule
%   $\MSD(\hat{\mu}_{cal})$ & 2.5 & 3.9 & 2.4 & 8.7 & 8.4 & 6.1\\ 
%   $\MSD(\hat{\mu}_{CLW})$ & - & - & - & 9.1 & 9.2 & 6.5\\ 
%    \bottomrule
% \end{tabular}
% \end{table}

\begin{table}[!h]
\centering
{
\caption{{MSD values (scaled to $10^2$)} of two population mean estimators, $\hat{\mu}_{cal}$ and $\hat{\mu}_{CLW}$, calculated using $12$ benchmark questions from $6$ surveys. Refer to Table \ref{est_list} for definition/formula of estimators and MSD definition in \eqref{eq:MSE}.} 
\label{tab:MSE_all}
\begin{tabular}{lrrrrrr}
  \hline
 Estimator & $P_1$ & $P_2$ & $P_3$ & $O_1$ & $O_2$ & $O_3$ \\ 
  \hline
  $\MSD(\hat{\mu}_{cal})$ & 0.080 & 0.210 & 0.097 & 1.039 & 1.024 & 0.480 \\
  $\MSD(\hat{\mu}_{CLW})$ & - & - & - & 1.168 & 1.237 & 0.564 \\ 
   \hline
\end{tabular}
}
\end{table}

In Table \ref{tab:MSE_all}, we compare $\hat{\mu}_{cal}$ and $\hat{\mu}_{CLW}$, in terms of MSD, provided in $10^2$ scale, for ease of representation. %{$\MSD(\hat{\mu}_{cal})$ with $\MSD(\hat{\mu}_{CLW})$ 
%(in $10^2$ or \% scale)
%, calculated using $12$ benchmark questions from $6$ surveys. 
{We see that for opt-in surveys, $\MSD(\hat{\mu}_{CLW})$ is greater than $\MSD(\hat{\mu}_{cal})$. For example, for $O_1$, $\MSD(\hat{\mu}_{CLW})$ is $1.168$, whereas $\MSD(\hat{\mu}_{cal})$ is $1.039$. Hence, no improvement is observed at an overall survey-level in opt-in surveys, using IPW as compared to calibrated weights.} As a result, in comparison to ABS surveys, MSD values for opt-in surveys are still quite high. For example, for $O_1$, {$\MSD(\hat{\mu}_{CLW})$} is $1.168$ whereas for $P_1$, {$\MSD(\hat{\mu}_{cal})$} is $0.080$. {Next, in Table \ref{tab:MSE_sub}, we compare MSD values of the same estimators at subgroup levels of age and race. Herein we observe that excepting a few cases, $\MSD(\hat{\mu}_{CLW})$ is greater than $\MSD(\hat{\mu}_{cal})$. The include $6$ cases out of $18$ as follows -- 
age group $18-29$ years for $O_1$, age group $65+$ years for $O_2$ and $O_3$, 
race White for $O_1$, 
race Black for $O_2$ and $O_3$. 
Finally, a similar conclusion as in Table \ref{tab:MSE_all} can be drawn that IPW doesn't improve upon the estimator using calibrated weights. Again, comparing opt-in surveys to ABS,} we observe that bias pertaining to specific subgroups is present in opt-in samples -- as reported by Pew. For example, for $O_1$ in age group $18-29$ years, {$\MSD(\hat{\mu}_{CLW})$} is $2.312$, as compared to the value $0.158$ of {$\MSD(\hat{\mu}_{cal})$} for the same category in $P_1$.

% \begin{table}[ht]
% \centering
% \caption{Subgroup level \AB{MSE} values (in \%) of $\hat{\mu}_{cal}$ and $\hat{\mu}_{CLW}$, calculated using $12$ benchmark questions from $6$ surveys at $3$ age subgroups ($18-29, 30-64, 65+$ years) and $3$ race subgroups (White, Black and Hispanic). Estimates for White and Black adults are based on those who do not identify as Hispanic and values for Asian/other race categories (constituting less than 5\% respondents in each survey) are not reported.} 
% \label{tab:MSE_sub}
% \begin{tabular}{lllllllll}
%   \toprule
%     & Factor & Group & $P_1$ & $P_2$ & $P_3$ & $O_1$ & $O_2$ & $O_3$ \\ 
%   \midrule
%   & & 18-29 (yrs) & 2.9 & 6.6 & 3.2 & 15 & 15.7 & 11.8\\
%   \AB{$\MSD(\hat{\mu}_{cal})$} & age & 30-64 (yrs) & 2.7 & 3.9 & 2.8 & 9.9 & 10.2 & 6.6 \\ 
%   & & 65+ (yrs) & 2.7 & 3.4 & 2.7 & 3.4 & 3.2 & 2.3 \\ 
%   \midrule
%   & & 18-29 (yrs) & - & - & - & 13.2 & 17.1 & 11.6 \\ 
%   \AB{$\MSD(\hat{\mu}_{CLW})$} & age & 30-64 (yrs) & - & - & - & 10.6 & 11.2 & 7.4 \\ 
%   & & 65+ (yrs) & - & - & - & 4.3 & 3.0 & 2.3 \\
%  \midrule 
%   & & White & 2.3 & 3.3 & 2.5 & 8.1 & 8.1 & 5.2\\ 
%   \AB{$\MSD(\hat{\mu}_{cal})$} & race & Black & 4.1 & 5.2 & 4.2 & 8.4 & 8.9 & 6.1\\ 
%   & & Hispanic & 2.7 & 5.3 & 4 & 13.8 & 13.1 & 13 \\ 
%    \midrule
%   & & White  & - & - & - & 7.9 & 9.2 & 5.8 \\ 
%   \AB{$\MSD(\hat{\mu}_{CLW})$} & race & Black & - & - & - & 9.1 & 7.2 & 5.4 \\ 
%   & & Hispanic  & - & - & - & 16.6 & 14.3 & 14.6 \\ 
%    % \hline
%    \bottomrule
% \end{tabular}
% \end{table}

\begin{table}[!htb]
\centering
{
\caption{Subgroup level {MSD  values (scaled to $10^2$)} of $\hat{\mu}_{cal}$ and $\hat{\mu}_{CLW}$, 
%calculated using $12$ benchmark questions from $6$ surveys 
at $3$ age subgroups ($18-29, 30-64, 65+$ years) and $3$ race subgroups (White, Black and Hispanic). Estimates for White and Black adults are based on those who do not identify as Hispanic and values for Asian/other race categories (constituting less than 5\% respondents in each survey) are not reported.} 
\label{tab:MSE_sub}
\begin{tabular}{lllllllll}
  \toprule
    & Factor & Group & $P_1$ & $P_2$ & $P_3$ & $O_1$ & $O_2$ & $O_3$ \\ 
  \midrule
  & & 18-29 (yrs) & 0.158 & 0.507 & 0.177 & 2.869 & 3.162 & 1.790 \\ 
  $\MSD(\hat{\mu}_{cal})$ & age & 30-64 (yrs) & 0.104 & 0.239 & 0.117 & 1.345 & 1.407 & 0.567 \\
  & & 65+ (yrs) & 0.130 & 0.185 & 0.135 & 0.189 & 0.162 & 0.102 \\ 
  \midrule
  & & 18-29 (yrs) & - & - & - & 2.312 & 3.901 & 1.930 \\ 
  $\MSD(\hat{\mu}_{CLW})$ & age & 30-64 (yrs) & - & - & - & 1.497 & 1.705 & 0.728 \\
  & & 65+ (yrs) & - & - & - & 0.318 & 0.146 & 0.097 \\ 
 \midrule 
  & & White & 0.064 & 0.152 & 0.09 & 1.034 & 0.907 & 0.39 \\ 
  $\MSD(\hat{\mu}_{cal})$ & race & Black & 0.281 & 0.502 & 0.376 & 1.099 & 1.121 & 0.499 \\ 
  & & Hispanic & 0.102 & 0.375 & 0.236 & 2.625 & 2.341 & 2.147 \\ 
   \midrule
  & & White  & - & - & - & 0.913 & 1.136 & 0.471 \\ 
  $\MSD(\hat{\mu}_{CLW})$ & race & Black & - & - & - & 1.351 & 0.877 & 0.464 \\ 
  & & Hispanic  & - & - & - & 3.441 & 2.781 & 2.444 \\
   \bottomrule
\end{tabular}
}
\end{table}

\begin{figure}[!htb]
\caption{Plot of squared differences of two population mean estimates with benchmark $\bar{Y}$, i.e., $(\bar{Y} -\hat{\mu}^{P_3}_{cal})^2$ in black and $(\bar{Y} - \hat{\mu}^{O_3}_{CLW})^2$ in blue, for $12$ questions and $3$ age groups $18-29, 30-64, 65+$ years. Highest values of blue bars in each age group are denoted in dotted lines. $Y$-axis is scaled to $10^3$.}
\label{fig:ABS_CLW}
\centering
  \includegraphics[width=0.77\linewidth]{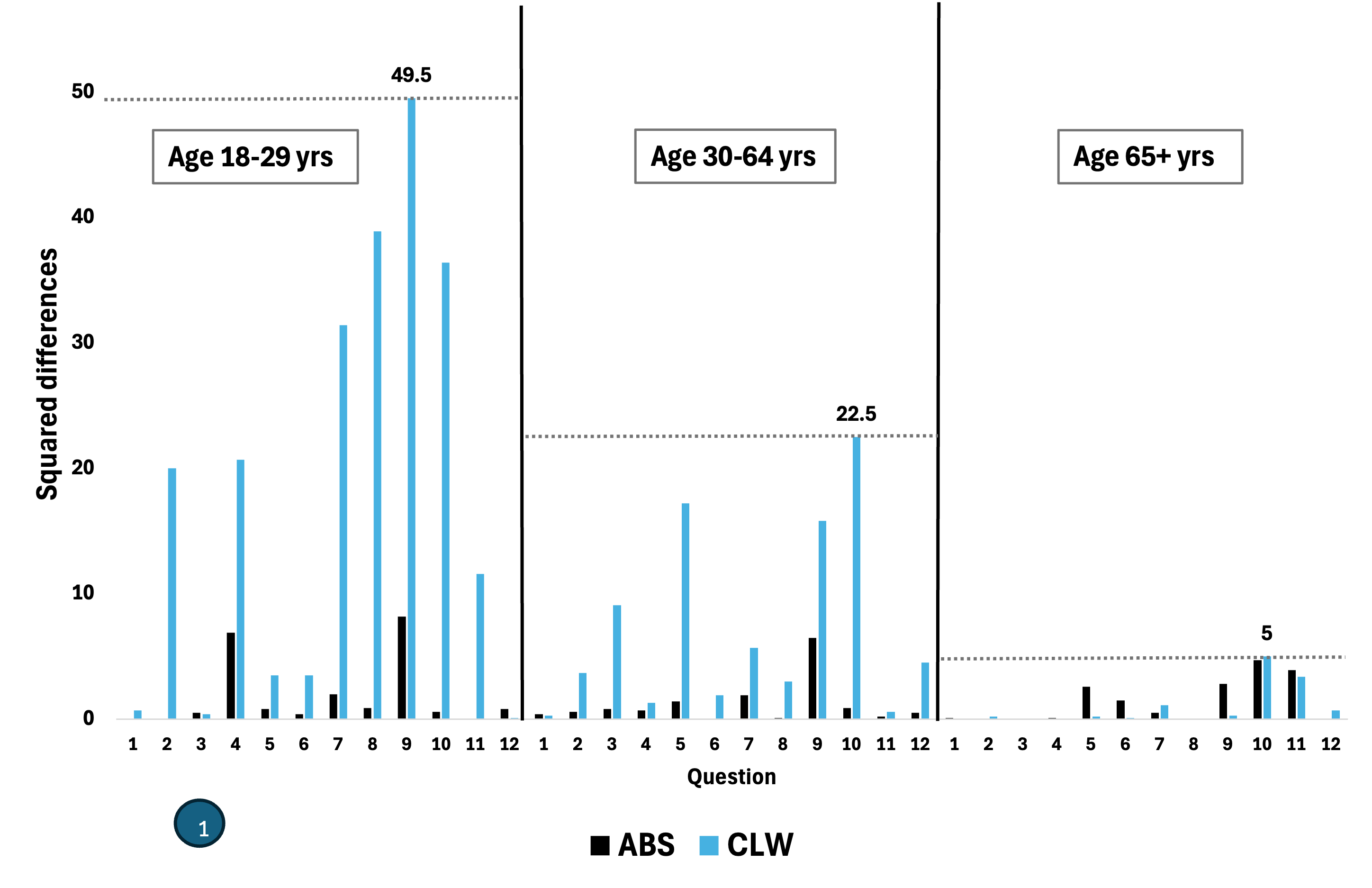}
\end{figure}

To understand the impact of IPW method on selection bias improvement at a more granular level, we calculate the {squared} difference of estimators ($\hat{\mu}_{CLW}$ and $\hat{\mu}_{{cal}}$) from benchmark values at $3$ subgroup levels of age for $12$ questions. In this section, we discuss the squared difference results for surveys $O_3$ and $P_3$ only, 
as in the next section we will use the other surveys ($O_1$, $O_2$ and $P_1$, $P_2$) to construct an estimator of bias. Hereon, we use superscripts on estimators to denote the source sample, such as $P_1,P_2,P_3$ or $O_1,O_2,O_3$.
In Figure \ref{fig:ABS_CLW}, light blue bars denote squared differences for $\hat{\mu}_{CLW}^{O_3}$ and black bars denote the same for $\hat{\mu}_{{cal}}^{P_3}$. We mark the highest value of blue bars for each age group in dotted line. The squared difference values in this figure and the rest of the figures in Section \ref{sec_analysis} are scaled to $10^3$, for ease of visualization.
Similar conclusions as before can be drawn here -- among the three age groups, group $18-29$ years displays the highest difference between heights of the two types of bars. 
We observe from the figure that the highest value of squared difference for $\hat{\mu}_{CLW}^{O_3}$ is $49.473$ in the age group $18-29$ years. {This value comes from Question 9 regarding Food Stamps. Bias is often significant in questions that are sensitive, subjective, or challenging to answer due to ambiguous definitions of concepts or reliance on the respondent's memory. Based on these criteria, estimates for Question 9 are more likely to have high squared differences from benchmark values.
Overall, we see that the blue bars are quite high, compared to the black bars.} Thus, we conclude that improving representativeness alone does not improve estimates from opt-in surveys.

\subsection{Measurement difference bias-corrected IPW estimators}
\label{sec_mebc}
In this section, we address bias due to measurement {differences}. {The findings in \cite{KML} suggest that all the opt-in samples suffer from a similar bias. The authors confirm that this measurement error bias is due to the effect of age and race, which can also be statistically tested using ANOVA models. Hence, following the empirical investigation of \cite{KML}, we assume that the distribution of bias is similar between the opt-in samples. With this prior knowledge of factors causing bias and the assumption of similar bias between opt-in samples}, we compute estimate of bias using $O_1, O_2$ and $P_1, P_2$, keeping $O_3$ for bias correction and $P_3$ for validation. 
Recall that from the model defined in \eqref{eq:model}, $\hat\mu_{CLW} - \hat \mu_{cal}$ yields an unbiased estimator of $\epsilon$. Thus, we define an estimate of the bias at question $\times$ age group level as the mean of pair-wise differences between $nps$ and $ps$. For question $j$ and age group level $c$ , this is given by
\begin{align} \label{eq:eps}
    \hat{\epsilon}^{jc} & \equiv {4}^{-1} \Big \lbrace (\hat{\mu}_{CLW}^{jc;O_1} - \hat{\mu}_{{cal}}^{jc;P_1}) + (\hat{\mu}_{CLW}^{jc;O_1} - \hat{\mu}_{{cal}}^{jc;P_2}) + (\hat{\mu}_{CLW}^{jc;O_2} - \hat{\mu}_{{cal}}^{jc;P_1})+ (\hat{\mu}_{CLW}^{jc;O_2} - \hat{\mu}_{{cal}}^{jc;P_2}) \Big \rbrace \notag\\
    & = {2}^{-1} \Big \lbrace (\hat{\mu}_{CLW}^{jc;O_1} - \hat{\mu}_{{cal}}^{jc;P_1}) + (\hat{\mu}_{CLW}^{jc;O_2} - \hat{\mu}_{{cal}}^{jc;P_2}) \Big \rbrace,
\end{align}
where $j = 1,\cdots,12; \; c = 1,2,3$. 
Due to the additive bias structure in the model defined in \eqref{eq:model},
we calculate the bias-corrected estimates at question $j$ and age group level $c$ in $O_3$ as
\begin{align} \label{eq:bc}
    \hat{\mu}^{jc;O_3}_{bc;CLW} \equiv \hat{\mu}^{jc;O_3}_{CLW} - {\hat{\epsilon}^{jc}}, \quad j = 1,\cdots,12;\;\; c = 1,2,3.
\end{align}
In this section, we describe the bias correction method focusing on the factor age, but similar correction could be done using the factor race. We do not consider both factors together (age $\times$ race) for the bias-correction, due to unavailability of benchmarks at such granular levels.

\begin{figure}[!htb]
\caption{Plot of {squared differences} of $\hat{\mu}^{O_3}_{CLW}$ and its bias-corrected version $\hat{\mu}^{O_3}_{bc;CLW}$ with benchmark, i.e., {$(\bar{Y} -\hat{\mu}^{O_3}_{CLW})^2$} in blue and {$(\bar{Y} - \hat{\mu}^{O_3}_{bc;CLW})^2$} in green, for $12$ questions and $3$ age groups $18-29, 30-64, 65+$ years. Highest values of green bars in each age group are denoted in dotted lines. $Y$-axis is scaled to $10^3$.} \label{fig:CLWbc}
\centering
  \includegraphics[width=0.77\linewidth]{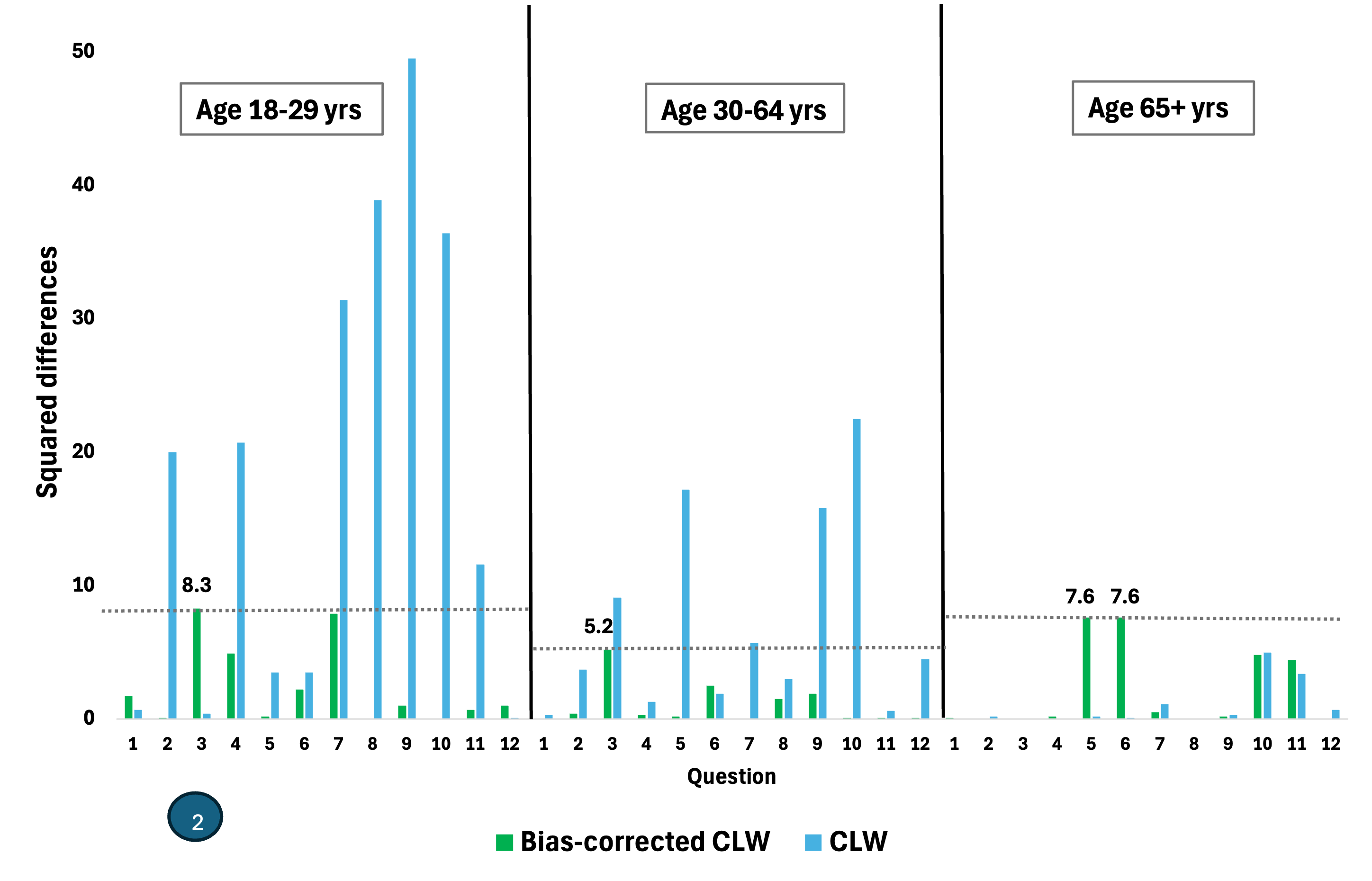}
\end{figure}

We compare $\hat{\mu}^{jc;O_3}_{bc;CLW}$ with $\hat{\mu}^{jc;O_3}_{CLW}$ in terms of {squared differences} with benchmarks to see the effect of measurement bias correction due to age. In Figure \ref{fig:CLWbc}, maintaining color consistency with the previous figure, we plot the {squared differences} of $\hat{\mu}^{jc;O_3}_{CLW}$ in blue bars and that of bias-corrected estimates $\hat{\mu}^{jc;O_3}_{bc;CLW}$ in green bars. The highest value of green bars in every age group is marked in dotted line. We see that in most cases, i.e. question $\times$ age group cell, the green bars are lower in height than blue bars. The highest value of green bar in age group $18-29$ years is $8.315$. This is much lower as compared to the highest value of blue bar in the same group ($49.473$ as noted in Figure \ref{fig:ABS_CLW}). Thus, from the empirical investigation we conclude that improving measurement {differences} after improving representativeness produces better estimates.

\subsection{Composite estimator from ABS and opt-in samples}
\label{sec:comp_est}
\begin{figure}[!htb]
\caption{Plot of {squared differences} with benchmark of $\hat{\mu}^{O_3}_{bc:CLW}$ (using $O_3$) and {$\tilde\mu_{comb}$} (using $O_3$ and $P_3$), i.e., {$(\bar{Y} -\hat{\mu}^{O_3}_{bc:CLW})^2$} in green and {$(\bar{Y} - \tilde{\mu}_{comb})^2$} in orange, for $12$ questions and $3$ age groups $18-29, 30-64, 65+$ years. Highest values of orange bars in each age group are denoted in dotted lines. $Y$-axis is scaled to $10^3$.} \label{fig:comb}
\centering
  \includegraphics[width=0.77\linewidth]{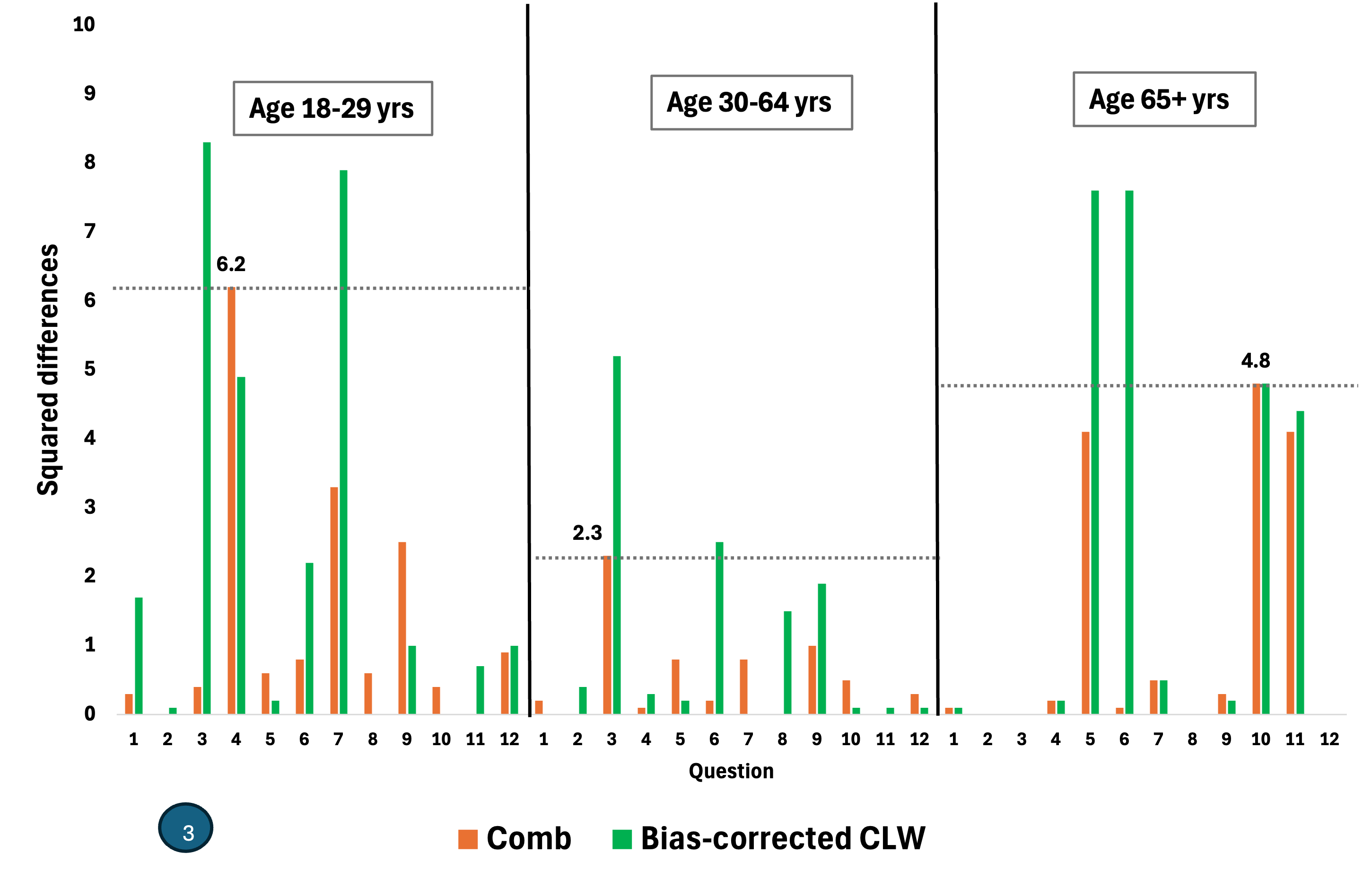}
\end{figure}

\begin{figure}[!htb]
\caption{Plot of squared differences with benchmark of composite estimators $\hat{\mu}_{EV}$ and {$\tilde\mu_{comb}$}, i.e., {$(\bar{Y} -\hat{\mu}_{EV})^2$} in dark blue and {$(\bar{Y} - \tilde{\mu}_{comb})^2$} in orange, for $12$ questions and $3$ age groups $18-29, 30-64, 65+$ years. Highest values of orange bars in each age group are denoted in dotted lines. $Y$-axis is scaled to $10^3$.} \label{fig:EV}
\centering
  \includegraphics[width=0.77\linewidth]{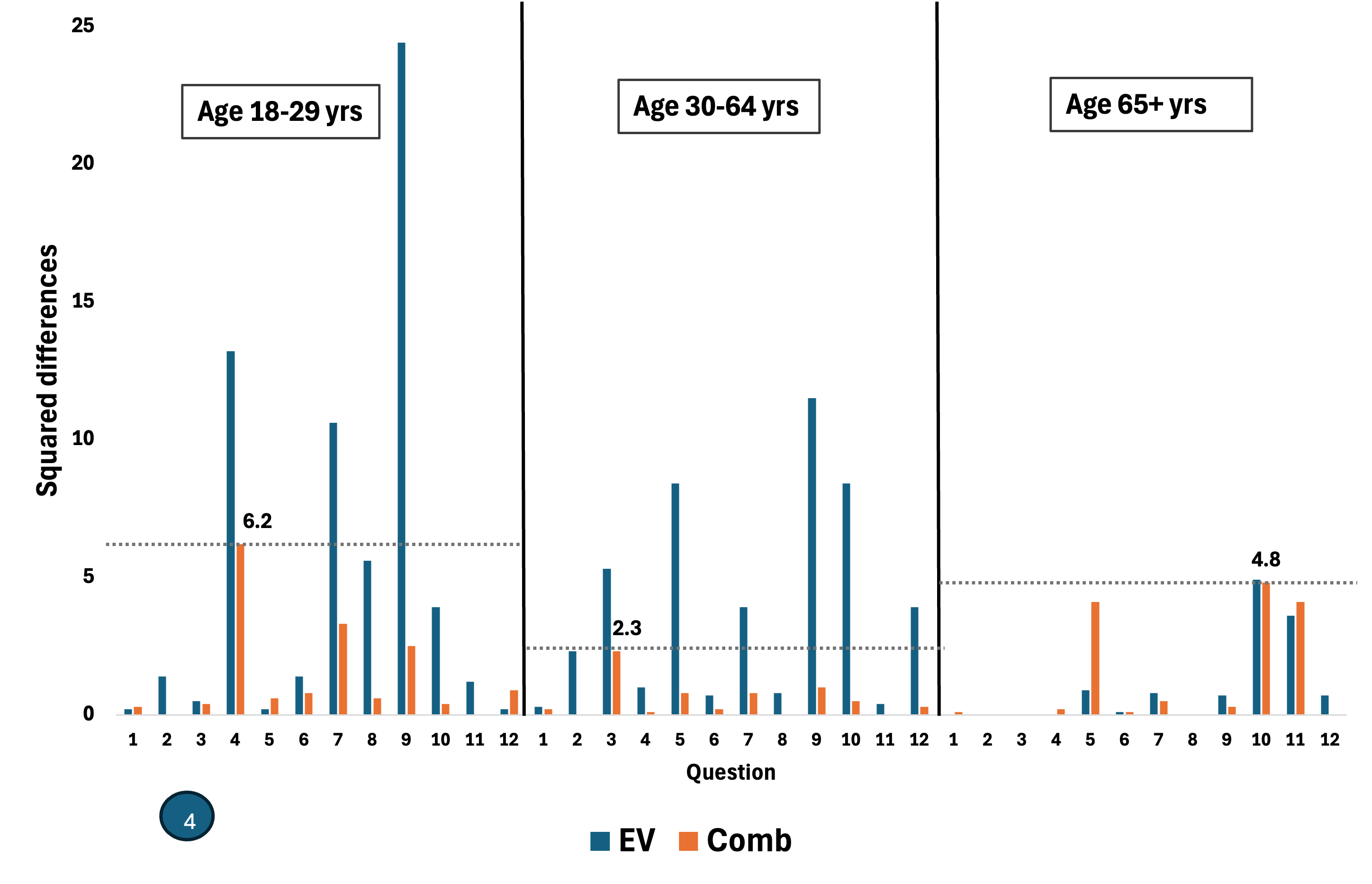}
\end{figure}

We calculate the composite estimators $\hat{\mu}_{EV}$ and {$\tilde\mu_{comb}$} described in Section \ref{sec_comp}. For a fair comparison, for both the estimators we take the estimate of bias $\hat{\epsilon}$ %(in model \eqref{eq:model}) 
as defined in \eqref{eq:eps} in Section \ref{sec_mebc}, and the values of $v_1, v_2$ as the estimated variances of $\hat{\mu}_{CLW}$ and $\hat{\mu}_{{cal}}$ (from \texttt{R} packages \texttt{nonprobsvy} and \texttt{survey} respectively). {We show two figures in this section -- Figure \ref{fig:comb} and \ref{fig:EV}. In the first one we compare proposed composite estimator ({$\tilde\mu_{comb}$}) with the bias-corrected estimator from the previous section ($\hat{\mu}^{O_3}_{bc:CLW}$). In the second one we compare our proposed composite estimator ({$\tilde\mu_{comb}$}) with the existing composite estimator $\hat{\mu}_{EV}$.}
In Figure \ref{fig:comb}, we plot the {squared} differences of {$\tilde\mu_{comb}$} from benchmark in orange bars and that of $\hat{\mu}^{O_3}_{bc:CLW}$ in green bars (as in Figure \ref{fig:CLWbc}). 
%{We have reduced the scale of the vertical axis, for better visualization purposes.}
In this plot we see that on an average the orange bars are lower in height than the green bars, implying that the proposed composite estimator is even better than bias-corrected IPW estimator. The highest value of orange bar in age group $18-29$ years is $6.182$. This is lower in comparison to the same for green bar, which is $8.315$ (as shown in Figure \ref{fig:CLWbc}).

In Figure \ref{fig:EV}, we similarly compare {$\tilde\mu_{comb}$} and $\hat{\mu}_{EV}$ in terms of their {squared} differences with benchmarks. Again, on an average, the values for {$\tilde\mu_{comb}$} (denoted in orange bars) are lower than that of $\hat{\mu}_{EV}$ (denoted in dark blue bars). We observe from the dotted lines that the highest {squared} difference with benchmark is also lower for {$\tilde\mu_{comb}$} than $\hat{\mu}_{EV}$ in age groups $18-29$ and $30-64$ years. However, in the $65+$ age group, the performance of {$\tilde\mu_{comb}$} and $\hat{\mu}_{EV}$ is similar. Hence, we conclude that our proposed composite estimator performs better than the earlier one in the literature.

\subsection{Comparison of composite estimator with ABS}

\begin{figure}[!htb]
\caption{Plot of {squared differences} of $\hat{\mu}^{P_3}_{{cal}}$ (using $P_3$) and {$\tilde\mu_{comb}$} (using $P_3$ and $O_3$) with benchmark, i.e., {$(\bar{Y} -\hat{\mu}^{P_3}_{cal})^2$} in black and {$(\bar{Y} - \tilde{\mu}_{comb})^2$} in orange, for $12$ questions and $3$ age groups $18-29, 30-64, 65+$ years. Highest values of orange bars in each age group are denoted in dotted lines. $Y$-axis is scaled to $10^3$.} \label{fig:ABS_comb}
\centering
  \includegraphics[width=0.77\linewidth]{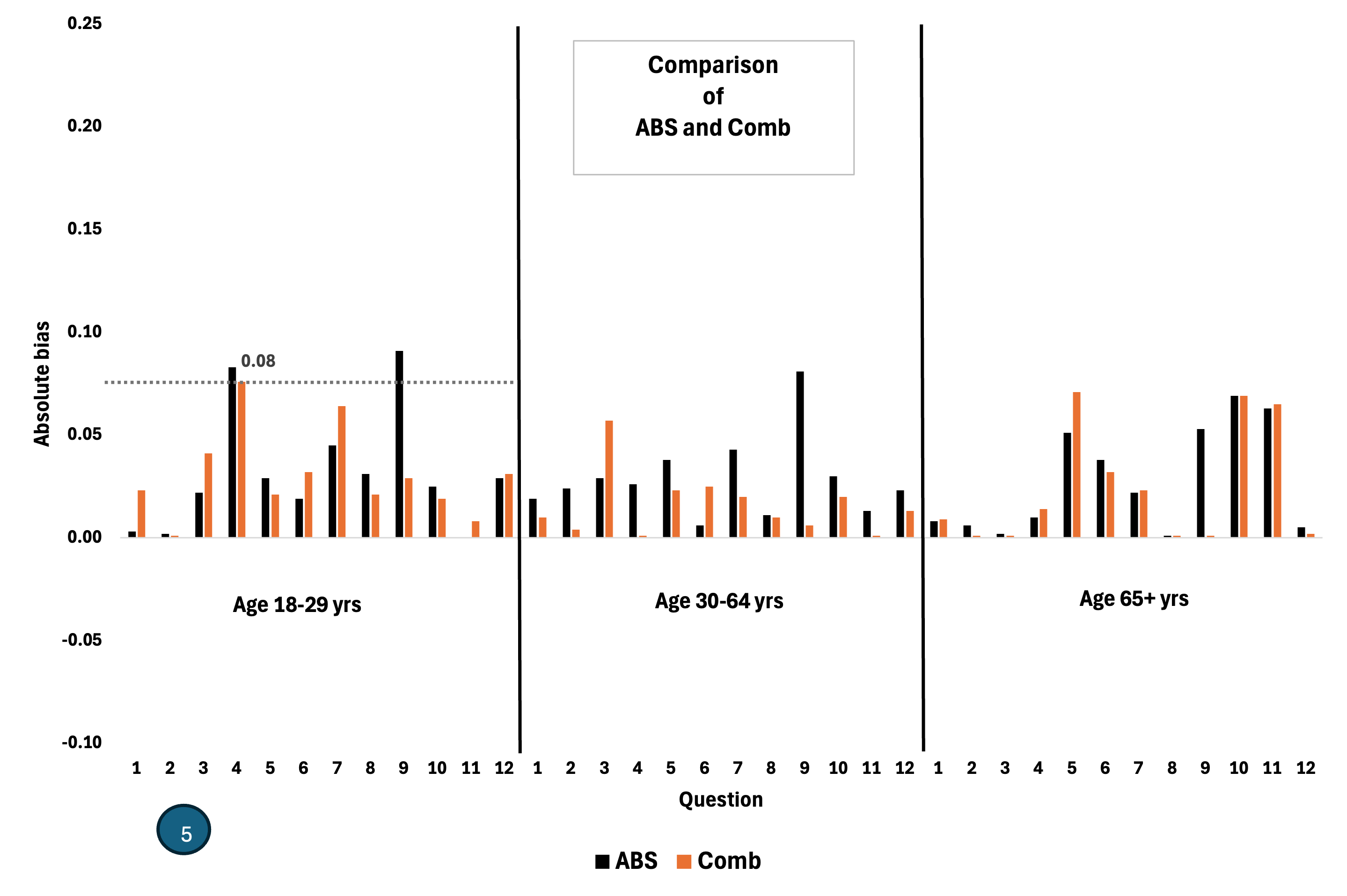}
\end{figure}

After the discussion about composite estimators in Section \ref{sec:comp_est}, one might be tempted to ask the question whether the proposed composite estimator (from both $ps$ and $nps$) outperforms the estimator from $ps$ alone. In this section, we want to investigate if the performance of composite estimator can be further improved. In Figure \ref{fig:ABS_comb}, we provide a comparison of {squared differences of $\tilde{\mu}_{comb}$} (denoted in orange bars) and $\hat{\mu}^{P_3}_{{cal}}$ (denoted in black bars), maintaining the color consistency from previous figures. For the age group $18-29$ years, we see that the estimator from $ps$ alone is better than our proposed composite estimator for some questions (viz. 1, 6, 7, 12). 
We also observe that for Question 10 (concerning social security) in the $65+$ age group, all estimators discussed in Section \ref{sec_analysis} exhibit similar values for squared differences from the benchmark (Figures \ref{fig:ABS_CLW}–\ref{fig:ABS_comb}).
This can be attributed to the fact that older or retired people are more likely to receive such benefits and are less susceptible to bogus responding. As a result, estimates from $nps$ are already close to $ps$.

In our case study $ps$ already has a large sample size (close to five thousand). As a result $ps$ itself is good enough (in most cases) for estimation of finite population characteristics. So in order to explain situations where performance of the proposed composite estimator is significantly better in terms of MSD , we generate $ps$ of smaller sample sizes. We do this by drawing samples from $P_3$ using a stratified sampling procedure with proportional allocation, where the strata are created by segmentation based on survey weights in $P_3$, viz., $(0,0.5), [0.5,1)$ and so on. We then calculate MSD values of $\hat{\mu}_{EV}$, $\tilde{\mu}_{comb}$ and $\hat{\mu}_{{cal}}$ for the smaller samples from $P_3$, keeping the original $nps$ $O_3$ intact. We note here that the estimators $\hat \mu_{CLW}$ and $\hat{\mu}_{bc;CLW}$ (computed using only $O_3$) remain the same, as there is no change in sample size of $O_3$. Their respective MSD values are {$0.564$} and $0.183$. In Table \ref{tab:ASD}, we provide MSD values (scaled to $10^2$, comparable to Table \ref{tab:MSE_all}), along with relative change in MSD (in \% inside parenthesis) of estimators with respect to $\hat{\mu}_{{cal}}$, defined as
\begin{align} \label{rel_per}
\frac{\MSD(\hat{\mu}_{cal}) - \MSD(\hat{\mu})}{\MSD(\hat{\mu})} \times 100,
\end{align}
where $\hat\mu \in \{\hat \mu_{EV}, \; \tilde{\mu}_{comb}\}$.

\begin{table}[ht]
\centering
\caption{
{MSD values (in $10^2$ scale)} of estimator $\hat{\mu}_{{cal}}$ (from $P_3$ and its samples), $\tilde{\mu}_{comb}$ and $\hat{\mu}_{EV}$ (from $P_3$ and its samples, combined with $O_3$), 
%for varying sample sizes of $P_3$, 
starting from the original sample size ($4,912$) decreasing up to $100$. Numbers presented in ($\cdot$) correspond to the relative change in MSD (in \%) of the estimators compared to $\hat \mu_{cal}$, defined in \eqref{rel_per}.} \label{tab:ASD}
\centering
    \begin{tabular}{crcc}
    \toprule
    $ps$ sample size & $\hat{\mu}_{{cal}}$ & {$\hat \mu_{EV}$} & {$\tilde{\mu}_{comb}$}  \\
    \midrule
    4912 & 0.097 & 0.337 (-71.284) & 0.101 (-3.767) \\  
   1000 & 0.231 & 0.295 (-21.718) & 0.158 (45.896) \\
  500 & 0.399 & 0.385 (3.642) & 0.289 (38.293) \\ 
  100 & 1.073 & 0.748 (43.476) & 0.626 (71.399) \\ 
    \bottomrule
    \end{tabular}
    \end{table}   

{We observe that as sample size of $ps$ decreases, the proposed composite estimator {$\tilde{\mu}_{comb}$} 
has lower MSD values in comparison to $\hat{\mu}_{cal}$. For example, for $ps$ sample size $100$, $\MSD(\tilde{\mu}_{comb})$ is $0.626$, which is much lower in comparison to the value of $1.073$ of $\MSD(\hat{\mu}_{cal})$. This is also reflected in the relative change in MSD value between these two estimators ($71.399\%$). However, for the original sample $P_3$, we see that $\MSD(\Tilde{\mu}_{comb}) > \MSD(\hat{\mu}_{cal})$ and relative change in MSD is $-3.767\%$. 
Similar phenomenon is observed for $\hat{\mu}_{EV}$, where $\MSD(\hat{\mu}_{EV}) < \MSD(\hat{\mu}_{cal})$ for $ps$ sample sizes $500$ and $100$. It is interesting to note that the MSD of $\hat{\mu}_{EV}$ is always more than that of {$\tilde{\mu}_{comb}$}.}
Finally, we conclude that the proposed composite estimator is not better than the estimator from $ps$ alone, when $ps$ has an adequate sample size. But as the sample size of $ps$ decreases the variance of $\hat{\mu}_{{cal}}$ increases, as a result the composite estimator {$\tilde{\mu}_{comb}$} outperforms $\hat{\mu}_{{cal}}$.

% \begin{table}[ht]
% \caption{{MSD} of $\hat{\mu}_{{cal}}$ (from $P_3$ and its samples) and {$\tilde{\mu}_{comb}$} (from $P_3$ and it samples, combined with $O_3$), 
% %for varying sample sizes of $P_3$, 
% starting from the original sample size ($4,912$) decreasing up to $100$. {Numbers presented in ($\cdot$) correspond to the relative change (in \%) in MSD of the estimators compared to $\hat \mu_{cal}$.}} \label{tab:ASD}
% \centering
%     \begin{tabular}{ccccccc}
%     \toprule
%     % & & & MSD &\\
%     % \cmidrule(l){2-4}
%     $ps$ sample size & $\hat{\mu}_{{cal}}$ & {$\hat \mu_{CLW}$} & {$\hat \mu_{bc;CLW}$} & {$\hat \mu_{EV}$} &{$\tilde{\mu}_{comb}$}  \\
%     \midrule
%     4912 & 2.4 (4) & & & & 2.3 \\
%     1000 & 3.7  ({37}) & & & & 2.7\\
%     500  & 4.5 ({36}) & & & & 3.3 \\
%     100  & 8.1 (72) & & & & 4.7 \\
%     \bottomrule
%     \end{tabular}
%     \end{table}   

\subsection{Predictive model for bias correction}

\begin{figure}[!h]
     \centering
      \caption{Results of predictive modeling on probability sample $P_3$}
      \label{fig:GBM}
     \begin{subfigure}{0.55\textwidth}
         \centering
         \caption{Tuning parameters of GBM}
\includegraphics[width=\textwidth]{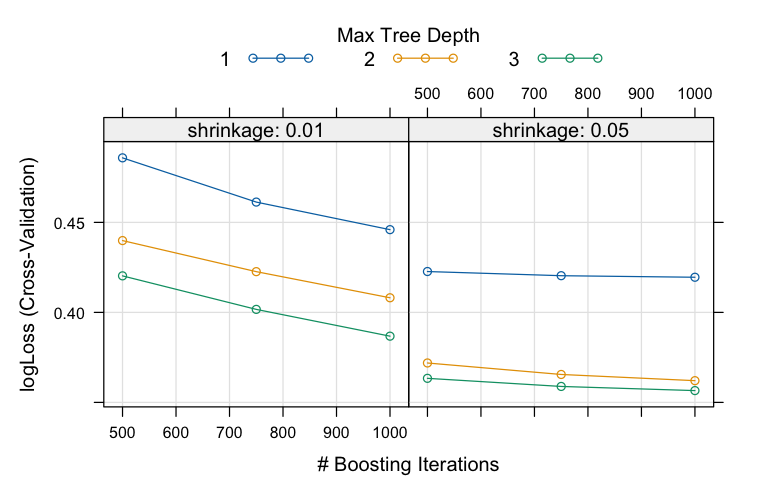}
     \end{subfigure}%
     \begin{subfigure}{0.5\textwidth}
         \centering
          \caption{ROC curve of RF and GBM}
    \includegraphics[width=\textwidth]{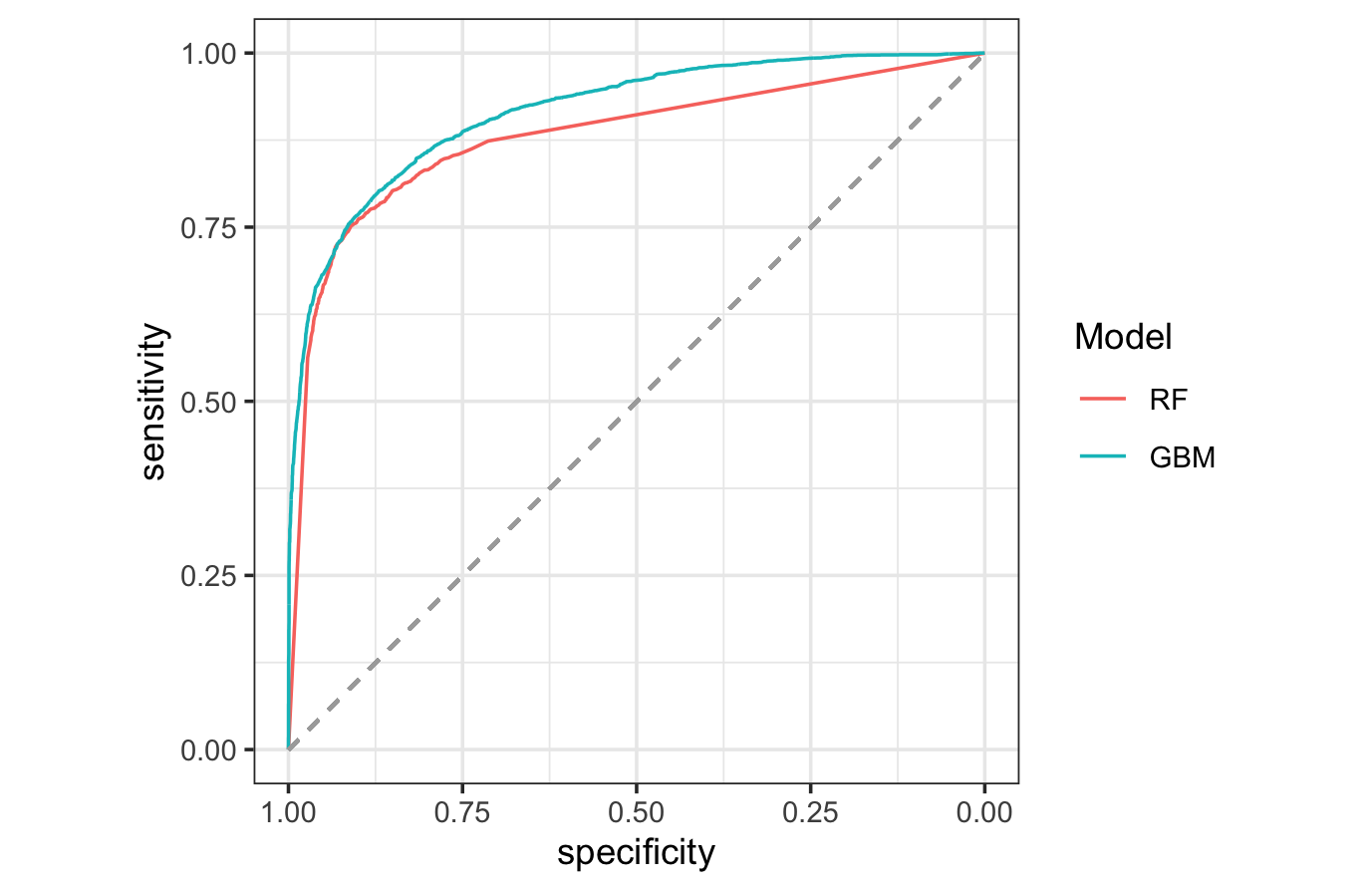}
     \end{subfigure}
     \label{sim-fig1}
\end{figure}

In Section \ref{sec_comp}, we mention that in order to compute composite estimators the bias $\epsilon$ needs to be known from alternate sources or historical data, which is why in Section \ref{sec_mebc} we estimate bias from $O_1, O_2, P_1, P_2$ and apply the bias correction in $O_3$. It is important to note that the estimate of bias proposed by \cite{Elliot} cannot be applied to our case, as then our composite estimator ends up being the estimator from $ps$ itself, i.e., in \eqref{eq:new-mu-comb-2}, using ${\hat\epsilon} = \bar{y}_2 - \bar{y}_1$, we get ${\tilde{\mu}_{comb}} = \bar{y}_1$. When other sources of information about bias are not available to obtain $\hat{\epsilon}$, we propose an alternative solution in this section. To this end, we fit a model on $P_3$ (relating responses to auxiliary variables), and using the fitted parameter estimates from this model, we predict the unit level responses in $O_3$ (or equivalently, the probabilities of responding `Yes'). For model building we use the \texttt{R} package \texttt{caret} (\cite{kuhn2020package}). With the help of \texttt{caret}, we fit machine learning (ML) models -- Random Forest (RF) and Gradient Boosting Model (GBM), using cross validation (CV) as the re-sampling method and tuning hyper-parameters such as number of boosting iterations, maximum tree depth, shrinkage, etc. In Table \ref{tab:mod_det} we provide necessary details on modeling building.
We observe from Table \ref{tab:mod_det} that both RF and GBM perform quite similarly in terms of accuracy ($0.85$) on the test data, but GBM has higher Area Under the Curve (AUC) value ($0.92$). In order to choose the final model, in addition to AUC and accuracy, we also leverage the {MSD} criteria considered in earlier sections. Finally, as GBM has lower {MSD} and higher AUC than RF, we choose this as our predictive model. The final values of tuning parameters used for the model are \texttt{n.trees} $=1000$, \texttt{interaction.depth} $=3$, \texttt{shrinkage} $=0.05$ and 
\texttt{n.minobsinnode} $=10$ and \texttt{logLoss} was used to select the optimal model using the smallest value. Question levels are among the topmost important variables responsible for classification. In Figure \ref{fig:GBM}, we show the \texttt{logLoss} with respect to different tuning parameters and also the Receiver Operating Characteristic (ROC) curves of the two binary classifier models (RF and GBM).

\begin{table}[!htb]
\centering
\caption{Details of the predictive modeling on $ps$ $P_3$ of $4912$ respondents, with the aim of obtaining bias-corrected estimates from $nps$ $O_3$.}
\label{tab:mod_det}
\begin{tabular}{ll}
\toprule
Data structure & Question $\times$ respondent level data \\
Size (Train $+$ test) & \begin{tabular}[c]{@{}c@{}} $59$ thousand observations split into \\ 
train ($80\%$) and test ($20\%$) \end{tabular} \\
Response variable type & Binary (Refusals are not considered) \\
Auxiliary variables (no. of levels) & \begin{tabular}[c]{@{}l@{}}Question ($12$), age ($3$), race-ethnicity ($5$), \\ education ($3$), gender ($3$), region ($4$) \end{tabular} \\
\begin{tabular}[c]{@{}l@{}} Predictive models \\ {(AUC, Accuracy)} \end{tabular} & \begin{tabular}[c]{@{}l@{}} %Logistic Regression ($80.9$), Decision Tree ($83.1$), \\
Random Forest (0.88, 0.85), \\
Gradient Boosting (0.92, 0.85)\end{tabular} \\
   \bottomrule
\end{tabular}
\end{table}

We then use the same IPW method (described in Section \ref{sec_rep}) on the predicted probabilities from the chosen model (GBM) to produce a new population mean estimator of $nps$. This is given by: 
\begin{align} \label{eq:IPW2}
  \hat{\mu}_{{ML};CLW} \equiv (\hat{N}^{A})^{-1} \sum_{i \in S_A} (\hat{p}^y_i/\hat{\pi}^A_i),  
\end{align}
where {$\hat{p}^y_i$ is the estimated likelihood that individual $i$ responds `Yes' to variable $y$ given the auxiliary characteristics of $i$, $\hat{\pi}^A_i$ is obtained from IPW method,} and $\hat{N}^A = \sum_{i \in S_A} (\hat{\pi}^A_i)^{-1}$. The difference of $\hat{\mu}_{{ML};CLW}$ defined in \eqref{eq:IPW2} with $\hat{\mu}_{CLW}$ defined in \eqref{eq:IPW} is that in \eqref{eq:IPW2} we use estimated {likelihood} values ($\hat{p}^y$) instead of true responses ($y$).

As a result, the new composite estimator in this case {is constructed as in \eqref{eq:comb}, but using \eqref{eq:IPW2} instead of \eqref{eq:IPW}}. The proposed estimator $\hat{\mu}_{{ML};comb}$ is defined as 
\begin{align}\label{eq:comb_m}
    \hat{\mu}_{{ML};comb} \equiv \Bigg( \frac{v_2}{v_1 + v_2} \Bigg) \hat{\mu}_{{cal}} + \Bigg ( \frac{v_1}{v_1 + v_2} \Bigg) \hat{\mu}_{{ML};CLW}.
\end{align}
We compare the above estimator with a {different version of $\hat{\mu}_{EV}$, which we denote by $\hat{\mu}_{{ML};EV}$. Although machine learning is not used in the construction of this estimator, for ease of comparison we use the same subscript `ML'. In this estimator, the unknown bias is estimated by the difference of weighted means from $ps$ and $nps$, as suggested by \cite{Elliot}.} Hence, $\hat{\mu}_{{ML};EV}$ is defined as
\begin{align}\label{eq:EVm}
    \hat{\mu}_{{ML};EV} \equiv \frac{\lbrace (\bar{y}_1 - \bar{y}_2)^2 + v_2 \rbrace \bar{y}_1 + v_1 \bar{y}_2}{(\bar{y}_1 - \bar{y}_2)^2 + v_1 + v_2}.
\end{align}

In Table \ref{tab:comp_new}, we show the values of the two composite estimators $\hat{\mu}_{{ML};comb}$ (proposed) and $\hat{\mu}_{{ML};EV}$ (existing) and their respective MSD values for $12$ questions considered for the analysis. Benchmark numbers are also provided alongside the estimates. We observe that for all questions, $\MSD(\hat{\mu}_{{ML};comb}) <\MSD(\hat{\mu}_{{ML};EV})$.

\begin{table}[ht]
\centering
\caption{Composite Estimators $\hat{\mu}_{{ML};comb}$ (proposed) and $\hat{\mu}_{{ML};EV}$ (existing) and their {MSD values (in $10^3$ scale)} for $12$ binary benchmark questions given in Table \ref{qs_det}.} \label{tab:comp_new}
\begin{tabular}{@{}lrcccc@{}}
  \toprule
 Question & Benchmark & $\hat{\mu}_{{ML};EV}$ & \MSD($\hat{\mu}_{{ML};EV}$) & $\hat{\mu}_{{ML};comb}$ & \MSD($\hat{\mu}_{{ML};comb}$) \\ 
  \midrule
1. Insurance & 90.800 & 88.979 & 4.450 & 90.755 & 4.447 \\ 
2. Blood Pressure & 31.100 & 37.725 & 5.271 & 35.709 & 5.109 \\
3. Parent & 26.000 & 22.336 & 10.189 & 25.977 & 10.099 \\ 
4. Food allergy & 9.400 & 14.038 & 0.984 & 12.823 & 0.966 \\ 
5. Job last year & 64.200 & 58.262 & 20.105 & 63.476 & 19.817 \\ 
6. Retirement account & 49.900 & 49.866 & 25.016 & 51.119 & 25.015 \\ 
\begin{tabular}[c]{@{}l@{}}7. Unemployment \\
compensation \end{tabular} & 9.300 & 16.846 & 0.970 & 12.189 & 0.825 \\
\begin{tabular}[c]{@{}l@{}} 8. Workers'\\
compensation \end{tabular} & 0.400 & 6.637 & 0.284 & 1.144 & 0.010 \\ 
9. Food stamps & 11.100 & 21.549 & 5.328 & 16.416 & 5.247 \\ 
10. Social Security & 21.800 & 29.898 & 26.574 & 26.189 & 26.413 \\
11. Union membership & 5.600 & 10.410 & 0.657 & 8.055 & 0.595 \\ 
12. U.S. citizenship  & 92.500 & 96.299 & 4.413 & 96.064 & 4.332 \\ 
   \bottomrule
\end{tabular}
\end{table} 

Finally, we summarize the two cases of bias-correction --
\begin{itemize}
    \item [\hypertarget{Case $1$:}{Case $1$:}] When bias is computed from alternative sources.
    \item [\hypertarget{Case $2$:}{Case $2$:}] When bias-corrected estimates are predicted using modeling. 
\end{itemize}
A comparison between the two cases would not be fair, given the fact that improved results are expected if bias is known from alternative sources. In Table \ref{tab:MAE_new}, we provide the {MSD} values of comparable estimators in these two scenarios. In \hyperlink{Case $1$:}{Case $1$:} we compare ${\tilde{\mu}_{comb}}, \; \hat{\mu}_{EV}, \;\hat{\mu}_{CLW}$ and $\hat{\mu}_{bc;CLW}$. Whereas in \hyperlink{Case $2$:}{Case $2$:} we compare $\hat{\mu}_{{ML};comb}, \;\hat{\mu}_{{ML};EV}$ and $\hat{\mu}_{{ML};CLW}$. We observe that in both cases the proposed estimators (${\tilde{\mu}_{comb}}$ and $\hat{\mu}_{{ML};comb}$) have lower {MSD} values than existing composite estimators ($\hat{\mu}_{EV}$ and $\hat{\mu}_{{ML};EV}$) as well as other estimators ($\hat{\mu}_{CLW}, \; \hat{\mu}_{bc;CLW}$ and $\hat{\mu}_{{ML};CLW}$). It is also noteworthy that the bias-correction is effective in both situations by significantly lowering {MSD} values. Referring back to Table \ref{tab:MSE_all} in Section \ref{sec:IPW_resutls}, we see that the MSD value of $\hat{\mu}_{cal}^{P_3}$ is $0.097$. Hence, the proposed composite estimators ($\tilde{\mu}_{comb}, \hat{\mu}_{{ML};comb}$) are doing a good job to achieve comparable accuracy for an $nps$ (combined with $ps$) as would be obtained normally for a $ps$ of similar size.

\begin{table}[ht]
\centering
{
\caption{Comparison of {MSD values (in $10^2$ scale, given inside parenthesis)} of proposed and competing estimators calculated using $12$ benchmark questions from $nps$ $O_3$ and $ps$ $P_3$ (in case of composite estimators) for two cases of bias correction. Refer to Table \ref{est_list} for details on the estimators.}
\label{tab:MAE_new}
\begin{tabular}{lccc}
  % \hline
  \toprule
   % & & Estimator (MSE) & \\
    % \cmidrule(l){2-4}
   & \begin{tabular}[c]{@{}c@{}}Proposed \\ Composite \end{tabular} & \begin{tabular}[c]{@{}c@{}}Existing \\ Composite \end{tabular} & \begin{tabular}[c]{@{}c@{}}Other\end{tabular}\\ 
  \midrule
 \begin{tabular}[c]{@{}l@{}} \hyperlink{Case $1$:}{Case $1$:} \end{tabular} & ${\tilde{\mu}_{comb} (0.101)}$ & $\hat{\mu}_{EV}(0.337)$ & \begin{tabular}[c]{@{}c@{}} $\hat{\mu}_{CLW}({0.564})$\\ $\hat{\mu}_{bc;CLW}(0.183)$ \end{tabular}\\
 \midrule
 \begin{tabular}[c]{@{}l@{}} \hyperlink{Case $2$:}{Case $2$:} \end{tabular} &  $\hat{\mu}_{{ML};comb}$(0.092) & $\hat{\mu}_{{ML};EV}$(0.355) & $\hat{\mu}_{{ML};CLW}$(0.099)\\
   \bottomrule
\end{tabular}
}
\end{table}

\section{Discussion}
\label{sec_conclusion}

In this article, we address the problem of improving finite population inference through  construction of composite estimators, integrating data from probability and nonprobability surveys. Our goal is to mitigate bias arising from issues of representativeness and measurement error in nonprobability surveys. To this end, we propose two bias correction methods -- 
%within a data integration framework --
(i) Measurement difference correction, applicable when additional data sources beyond a probability and a nonprobability survey are available; and
(ii) Model-based bias correction using advanced machine learning models, appropriate for settings where only a probability and a nonprobability survey are accessible. We illustrate the effectiveness of these approaches through a case study based on the data from \citet{KML}, showing that the proposed estimators yield improved accuracy relative to large government surveys regarded as gold standards. Method (i) offers a computationally straightforward solution, while method (ii) is more broadly applicable in practice, especially in scenarios with limited data sources.

We also demonstrate that the proposed composite estimator achieves improved performance (specifically, lower MSD) when the sample size of the probability survey is smaller relative to that of the nonprobability survey. However, reducing the probability sample size can give rise to small area estimation challenges. In our case study, we observe that some subgroups (defined by question × age group combinations) have no observations when the probability sample size is substantially reduced. In such situations, small area modeling techniques, as discussed in \citet{Nandram} and \citet{Franco}, become particularly valuable. Addressing this issue presents an interesting direction for future research.

The proposed methodology in this paper is tested on a specific case study involving opt-in surveys in the U.S. The performance of the method may vary across different countries, survey modes, or cultural contexts, where the nature of selection bias and measurement error may differ substantially. Additionally, the proposed methodology relies on certain assumptions, which are discussed in this paper along with references that address potential violations. As a direction for future research, it would be valuable to develop composite estimators that account for measurement errors in both probability and nonprobability surveys. Overall, our dual-focus approach of tackling both selection bias and measurement differences offers a promising framework for improving survey methodology in the context of growing reliance on nonprobability surveys.

\section{Acknowledgements}

The authors wish to thank Andrew Mercer, Senior Research Methodologist at the Pew Research Center, for his support on this research. The authors are also thankful to the two anonymous referees and associate editor for giving constructive comments on an earlier version of this article.

%%%%%%%%%%%%%%%%%%%%%%%%%%%%%%%%%%%%%%%%%%%%%%%%%%%%%%%%%%%%%%
\bibliographystyle{chicago}
\bibliography{ref.bib}

\end{document}